\documentclass[aps, prl, twocolumn, floatfix, superscriptaddress,nofootinbib,longbibliography]{revtex4-1} 
\usepackage{graphicx}
\usepackage{epstopdf}
\usepackage{bm, amsmath, amssymb}
\usepackage{float}
\usepackage{placeins}
\usepackage{color}

\newcommand{\ket}[1]{\left|#1\right>}
\newcommand{\bra}[1]{\left<#1\right|}

\begin{document}

\title{Driving forbidden transitions in the fluxonium artificial atom}

\author{U. Vool}
\altaffiliation{Current Address: Department of Physics, Harvard University, Cambridge, MA 02138 USA}
\email{uri\_vool@fas.harvard.edu}
\affiliation{Department of Applied Physics and Physics, Yale University, New Haven, CT 06520, USA}
\author{A. Kou}
\affiliation{Department of Applied Physics and Physics, Yale University, New Haven, CT 06520, USA}
\author{W. C. Smith}
\affiliation{Department of Applied Physics and Physics, Yale University, New Haven, CT 06520, USA}
\author{N. E. Frattini}
\affiliation{Department of Applied Physics and Physics, Yale University, New Haven, CT 06520, USA}
\author{K. Serniak}
\affiliation{Department of Applied Physics and Physics, Yale University, New Haven, CT 06520, USA}
\author{P. Reinhold}
\affiliation{Department of Applied Physics and Physics, Yale University, New Haven, CT 06520, USA}
\author{I. M. Pop}
\affiliation{Physikalisches Institut, Karlsruhe Institute of Technology, Karlsruhe 76131, Germany}
\author{S. Shankar}
\affiliation{Department of Applied Physics and Physics, Yale University, New Haven, CT 06520, USA}
\author{L. Frunzio}
\affiliation{Department of Applied Physics and Physics, Yale University, New Haven, CT 06520, USA}
\author{S. M. Girvin}
\affiliation{Department of Applied Physics and Physics, Yale University, New Haven, CT 06520, USA}
\author{M. H. Devoret}
\email{michel.devoret@yale.edu}
\affiliation{Department of Applied Physics and Physics, Yale University, New Haven, CT 06520, USA}

\date{\today}
\hyphenation{cir-cuit}

\begin{abstract}

Atomic systems display a rich variety of quantum dynamics due to the different possible symmetries obeyed by the atoms. These symmetries result in selection rules that have been essential for the quantum control of atomic systems. Superconducting artificial atoms are mainly governed by parity symmetry. Its corresponding selection rule limits the types of quantum systems that can be built using electromagnetic circuits at their optimal coherence operation points (``sweet spots"). Here, we use third-order nonlinear coupling between the artificial atom and its readout resonator to engineer the selection rules of our atom, allowing us to drive transitions forbidden by the parity selection rule for linear coupling to microwave radiation. A $\Lambda$-type system emerges from these newly accessible transitions, implemented here in the fluxonium artificial atom coupled to its ``antenna" resonator. We demonstrate coherent manipulation of the fluxonium artificial atom at its sweet spot by stimulated Raman transitions. This type of transition enables the creation of new quantum operations, such as the control and readout of physically-protected artificial atoms.

\end{abstract}

\maketitle
 
\section{I. Introduction}
Atoms exhibit complex level-transition structures, which are governed by the interactions between their components. Superconducting artificial atoms, however, presently have much simpler level-transition structures.
As superconducting circuits emerge as a leading platform to investigate quantum information and coherent quantum physics~\cite{devoret_superconducting_2013}, there is growing interest in engineering their selection rules~\cite{liu_optical_2005,deppe_two-photon_2008,harrabi_engineered_2009,forn-diaz_broken_2016,goetz_parity-engineered_2017,lin_protecting_2017} to implement a larger variety of quantum dynamics in artificial atoms. 

All superconducting circuits at their sweet spots - operating points where the circuits are insensitive to certain environmental noise mechanisms - obey parity symmetry. This symmetry forbids transitions between states of the same parity under a microwave drive~\cite{liu_optical_2005,blais_quantum-information_2007}. Away from the sweet spots, all transitions become allowed, but at the cost of lower qubit coherence~\cite{ithier_decoherence_2005}. The ability to drive such transitions can lead to the implementation of a new class of artificial atoms. Moreover, it is necessary for the control and measurement of physically-protected qubits~\cite{doucot_physical_2012,brooks_protected_2013,bell_protected_2014,kou_fluxonium-based_2017} - circuits which implement error correction at the hardware level, but whose inherent protection results in the quasi-impossibility of manipulating them directly. Is it possible then, in superconducting circuits, to break the parity selection rule while still operating at the sweet spot? 

In this article, we present a method for driving forbidden transitions in superconducting artificial atoms. Using nonlinear coupling between the atom and an ancilla resonator, we create an atom with engineered selection rules while maintaining the symmetry, and thus the coherence properties, of the sweet spot. We implement nonlinear coupling using a fluxonium artificial atom~\cite{manucharyan_fluxonium:_2009,manucharyan_superinductance_2011} inductively coupled to an ``antenna" resonator \cite{pop_coherent_2014,vool_non-poissonian_2014}. We demonstrate the creation of a $\Lambda$-type system in which the states of the fluxonium can be manipulated by a resonator excitation using the now-allowed transition. We then use this structure to cool the fluxonium atom to one of its two lowest energy eigenstates by resonator decay through spontaneous Raman scattering. Finally, we present coherent manipulation of the fluxonium through the $\Lambda$-type system by driving Rabi oscillations between the fluxonium ground and excited states, using stimulated Raman transitions through a virtual resonator excitation. 

\begin{figure} [!b]
 \includegraphics[angle = 0, width = \columnwidth]{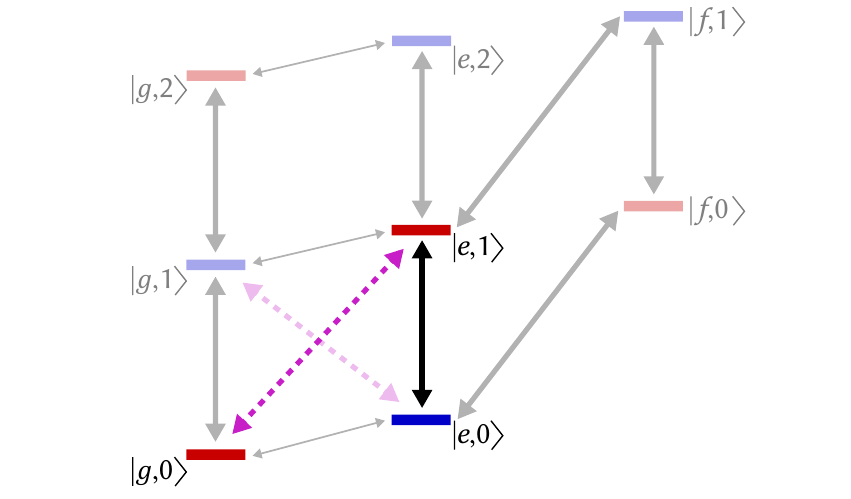}
 \caption{\label{fig1} Level diagram of a fluxonium atom at $\Phi^f_\mathrm{ext}/\Phi_0 = m + 1/2$ coupled to an electromagnetic resonator. At these sweet spots, only transitions between states of even parity (red) to states of odd parity (blue) are allowed. Transitions within the even or odd manifold are forbidden by a parity selection rule. By using nonlinear coupling between the resonator and the artificial atom, we can break the selection rule and drive forbidden transitions (magenta). We can thus construct a $\Lambda$-type system spanned by the $\ket{g,0}$, $\ket{e,0}$ and $\ket{e,1}$ states.
 }
 \end{figure}

\section{II. fluxonium-resonator system}
The fluxonium artificial atom is a superconducting circuit made up of a Josephson junction in parallel with a large linear inductance. While the fluxonium can be operated at any applied external flux through the fluxonium loop, $\Phi^f_\mathrm{ext}$, its flux-noise insensitive sweet spots are located at $\Phi^f_\mathrm{ext}/\Phi_0 = m,~m + 1/2$ where $\Phi_0$ is the magnetic flux quantum and $m \in \mathbb{N}$. Here, we will focus on the behavior of the fluxonium at its $m + 1/2$ sweet spots, where the transition frequency between its ground state $\ket{g}$ and first excited state $\ket{e}$ is $< 1~\mathrm{GHz}$. At the sweet spots, the potential of the fluxonium is a symmetric function of the flux $\bm{\phi_q}$ across the fluxonium junction. Hence, transitions are only allowed between states of opposite parity, and transitions such as $\ket{g} \leftrightarrow \ket{f}$ are forbidden by a parity selection rule~\cite{manucharyan_superinductance_2011,vool_engineering_2017}.

Coupling to a resonator breaks the fluxonium parity symmetry, but the total parity of the fluxonium-resonator excitations remains conserved and a parity selection rule still holds in the coupled system~\cite{blais_quantum-information_2007,liu_controllable_2006}. The level digram for this system, with its allowed and forbidden transitions, is shown in Fig.~\ref{fig1}. The states are labeled as $\ket{s,n}$ where $s$ is the state of the fluxonium and $n$ is the resonator photon number. States with even (odd) total parity are shown in red (blue). Transitions are only allowed between states of opposite total parity (solid black lines).The dashed magenta lines show forbidden transitions such as $\ket{g,0} \leftrightarrow \ket{e,1}$ and $\ket{g,0} \leftrightarrow \ket{e,1}$, and the ability to drive them gives us access to a $\Lambda$-type structure as shown in the figure. Low-frequency transitions such as $\ket{g,0} \leftrightarrow \ket{e,0}$ are not forbidden, but are suppressed due to a small dipole moment and the filtering of the fluxonium environment at low frequencies.

The parity selection rule only holds when the external drive is coupled to the odd $\bm{\phi_q}$ fluxonium operator via linear coupling to the antenna resonator. Nonlinear third-order coupling between the fluxonium and the resonator would lead to the drive coupling to operators such as $\bm{\phi_q}^2$ or $\bm{\phi_q}\bm{\phi_r}$, where $\bm{\phi_r}$ is the resonator flux. These even operators can drive transitions of equal parity at the fluxonium sweet spot (see Appendix B for more details). One would therefore like to implement third-order coupling between the fluxonium and its resonator to drive these forbidden transitions. Note that such coupling would preserve the protection offered at the sweet spot with respect to flux noise.

 \begin{figure} [!h]
\includegraphics[angle = 0, width = \columnwidth]{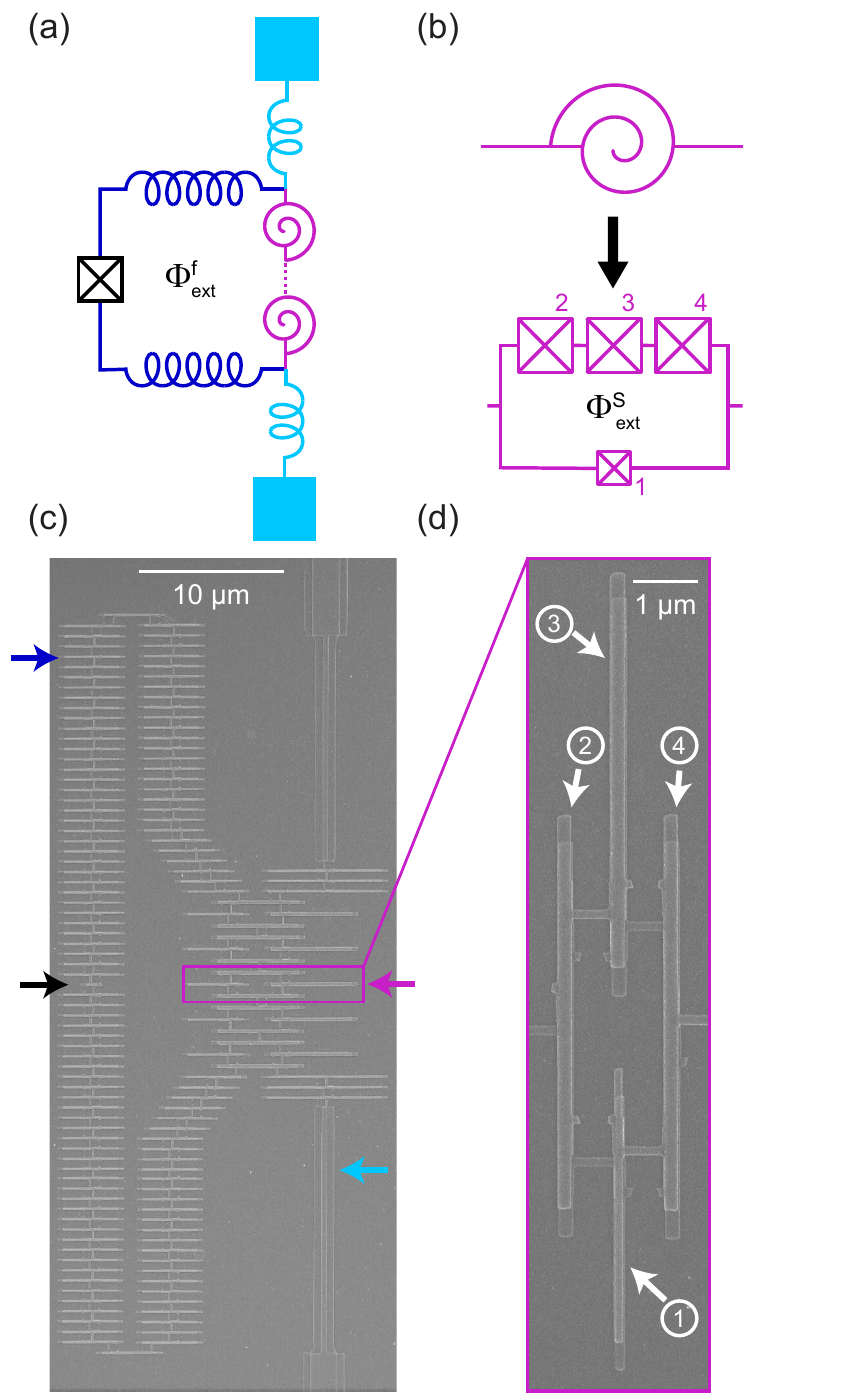}
\caption{\label{fig2} (a) Schematic of the fluxonium artificial atom coupled to an antenna resonator. The fluxonium is made up of a small phase-slip junction (black) shunted by a linear inductance (dark blue). It is coupled to the resonator (light blue) by sharing an inductance comprising SNAILS (magenta), which induce the nonlinear coupling. An external flux $\Phi^{f}_\mathrm{ext}$ is threaded through the fluxonium loop. (b) The SNAIL is composed of three large Josephson junctions in parallel with a smaller Josephson junction. An external magnetic flux $\Phi^{S}_\mathrm{ext}$ is threaded through the SNAIL loop. (c)  An SEM image of the device sketched in (a), with colored arrows indicating the different circuit elements. The area of the fluxonium loop is $A_f = 350\pm10 ~ \mathrm{\mu m^2}$ (d) An SEM image of a SNAIL, where the junctions corresponding to the numbered junctions in (b) are indicated. The ratio between the large junction and small junction areas is $\alpha = 0.4 \pm 0.02 $ and the area of the SNAIL loop is $A_S = 6\pm0.2 ~ \mathrm{\mu m^2}$. We can tune $\Phi^{f}_\mathrm{ext}$ and $\Phi^{S}_\mathrm{ext}$ quasi-independently using a global magnetic field due to the large ratio $A_f/A_S \simeq 60$.}
\end{figure}

\begin{figure*} [!t]

\includegraphics[angle = 0, width = \textwidth]{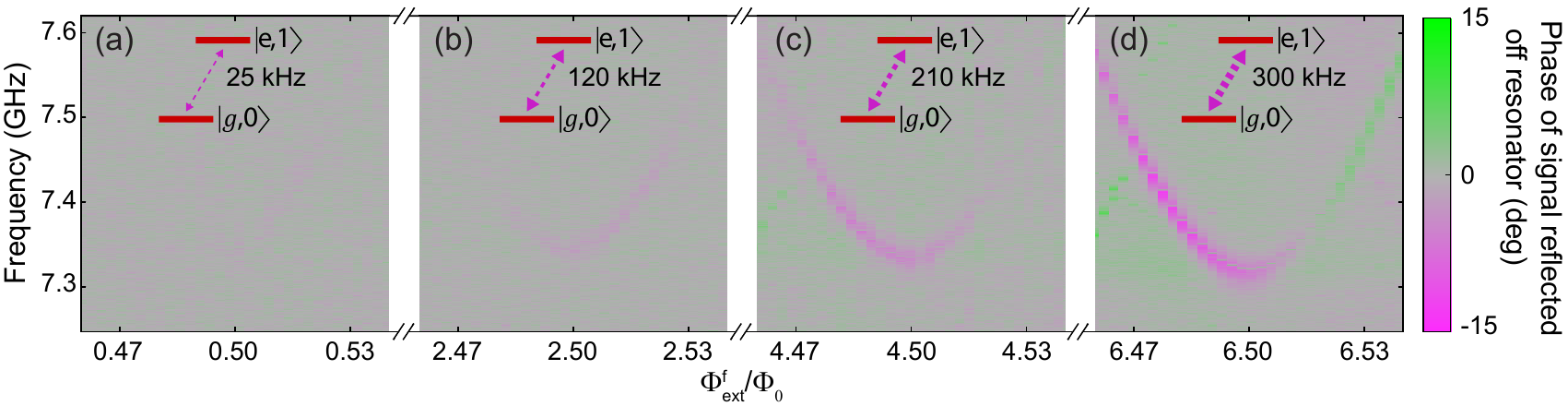}
\caption{\label{fig3}  (a)-(d) Two-tone spectroscopy of the $\ket{g,0}\leftrightarrow\ket{e,1}$ transition with a SNAIL-fluxonium device in the vicinity of $\Phi^{f}_\mathrm{ext}/\Phi_0 = 0.5$, $2.5$, $4.5$, and $6.5$, respectively. As the external flux is increased, the third-order coupling strength of the SNAIL grows, and the nominally forbidden transition becomes visible. The values for the $\ket{g,0}\leftrightarrow\ket{e,1}$ coupling strength are given in the insets. The change in the sign of the reflected phase near $\Phi^{f}_\mathrm{ext}/\Phi_0 = 6.51$ is due to a change in the dispersive coupling $\chi$ between the fluxonium and the resonator. 
}
\end{figure*}

Figure~\ref{fig2}a shows a diagram of a fluxonium atom nonlinearly coupled to its resonator. The fluxonium artificial atom is made up of a small junction (black) shunted by a large linear inductance (dark blue). The resonator is composed of a linear inductance and a capacitor (light blue). The nonlinear coupling is mediated by a nonlinear inductance made up of five three-wave-mixing dipole elements - each named the Superconducting Nonlinear Asymmetric Inductive eLement (SNAIL)~\cite{frattini_3-wave_2017} (magenta). 

A circuit diagram of the SNAIL is shown in Fig.~\ref{fig2}b, which consists of three large Josephson junctions in parallel with a smaller Josephson junction. As an external flux $\Phi^{S}_\mathrm{ext}$ is threaded through the SNAIL, it becomes a nonlinear element with third-order nonlinearity. The SNAIL design appears similar to that of the flux qubit~\cite{mooij_josephson_1999,wal_quantum_2000}, but the devices are operated in very different regimes. In the flux qubit, the area ratio ($\alpha$) between the small and large junction is chosen to be $ \approx 0.8$, which leads to a double-well potential at $\Phi^{S}_\mathrm{ext} = 0.5$. The SNAIL is designed to have $\alpha < 0.5$ to maintain a single potential well, and is operated around $\Phi^{S}_\mathrm{ext} = 0.1-0.4$ to create the asymmetric potential well necessary for three-wave mixing. The three-wave-mixing capability of the SNAIL has recently been proposed and implemented for quantum-limited amplification~\cite{zorin_josephson_2016,frattini_3-wave_2017,zorin_traveling-wave_2017}, but we use it here to implement nonlinear coupling between the fluxonium and its antenna resonator. Figure~\ref{fig2}d shows a scanning electron microscope (SEM) image of the SNAIL, composed of $\mathrm{Al}$/$\mathrm{AlOx}$/$\mathrm{Al}$ junctions, fabricated using the bridge-free-fabrication technique~\cite{lecocq_junction_2011,pop_fabrication_2012} on a sapphire substrate. In our implementation, $\alpha = 0.4 \pm 0.02$.

A SEM image of the full device is shown in Fig.~\ref{fig2}c. The large inductance of the fluxonium (dark blue) is built from 126 array junctions. The resonator inductance includes the five shared SNAILs as well as six unshared junctions. The antenna resonator capacitance is provided by two $0.5~\mathrm{mm}$ long leads. The resonator frequency was $f_r = 6.82~\mathrm{GHz}$ and the fluxonium qubit transition frequency at its $\Phi_0/2$ sweet spots was $f_q = 500~\mathrm{MHz}$. This sample was housed in a WR-102 waveguide and measured in reflection through an impedance-matched waveguide-SMA adapter~\cite{kou_simultaneous_2017}. 

\section{III. Spectroscopy}
In Fig.~\ref{fig3}, we show two-tone spectroscopy of the $\ket{g,0} \leftrightarrow \ket{e,1}$ transition around different fluxonium sweet spots. A continuous-wave tone was swept around the frequency of the $\ket{g,0} \leftrightarrow \ket{e,1}$ transition, while another tone at $f_r$ was used to measure the corresponding resonator response. The measurements were done around the $\Phi^f_\mathrm{ext}/\Phi_0 = m + 1/2$ fluxonium sweet spots, where $m=0,2,4,6$ for Fig.~\ref{fig3}a-d respectively. Though identical drive and measurement parameters were used at all the different sweet spots, the responses were significantly different. In Fig.~\ref{fig3}a, no transition is observed above the noise, which is consistent with previous fluxonium measurements of this forbidden transition~\cite{manucharyan_superinductance_2011}. In Figs.~\ref{fig3}b-d, however, the transition emerges and becomes increasingly visible at higher external flux. At higher external fluxes, the size of the three-wave-mixing term of the SNAIL increases, which results in a corresponding increase in the drive strength of the formerly forbidden transition. The insets in Fig.~\ref{fig3} show, for each sweet spot, the corresponding drive amplitude of the forbidden transition through the nonlinear coupling element. These values are based on a theoretical estimate of our three-wave-mixing coefficient, and a measurement of the drive amplitude from Raman experiments (see Appendices C \& D). The fluxonium coherence time at the $\Phi^f_\mathrm{ext}/\Phi_0 = 0.5$ and $6.5$ sweet spots was measured to be identical, and equal to $T_\mathrm{2R}=6~\mu s$. 

The frequency of the $\ket{g,0} \leftrightarrow \ket{e,1}$ transition is lowest at the sweet spots because it is the sum of the flux-dependent $\ket{g,0} \leftrightarrow \ket{e,0}$ fluxonium transition frequency and the approximately constant resonator frequency. The resonator frequency decreased slightly at higher external flux, due to the increasing linear inductance of the SNAIL. This manifests itself in our measurement as the minimum transition frequency being slightly lower for sweet spots of higher external flux.  Notice that in Fig.~\ref{fig3}d the sign of the phase response changes. This is likely due to a change in the fluxonium-resonator dispersive shift, which is also observable in a direct measurement of the fluxonium transition~\cite{vool_engineering_2017}.

\section{IV. $\Lambda$-system operations}
Driving parity-forbidden transitions, such as $\ket{g,0} \leftrightarrow \ket{e,1}$ and $\ket{e,0} \leftrightarrow \ket{g,1}$, allows us to construct a $\Lambda$-type system in which the fluxonium qubit states $\ket{g,0}$ and $\ket{e,0}$ are the low-energy states, and the excited state is a resonator excitation state such as $\ket{g,1}$ or $\ket{e,1}$. Previous superconducting implementations of $\Lambda$-type systems employed flux-tunable qubits away from their sweet spot~\cite{yang_quantum_2004,murali_probing_2004,valenzuela_microwave-induced_2006,grajcar_sisyphus_2008}, effective driven systems~\cite{inomata_microwave_2014,inomata_single_2016}, or two-photon transitions~\cite{kelly_direct_2010,novikov_raman_2016,earnest_realization_2017}. Here we present a physical implementation of $\Lambda$-type system at the fluxonium sweet spot using direct drives. 

\begin{figure} [t!]

\includegraphics[angle = 0, width = \columnwidth]{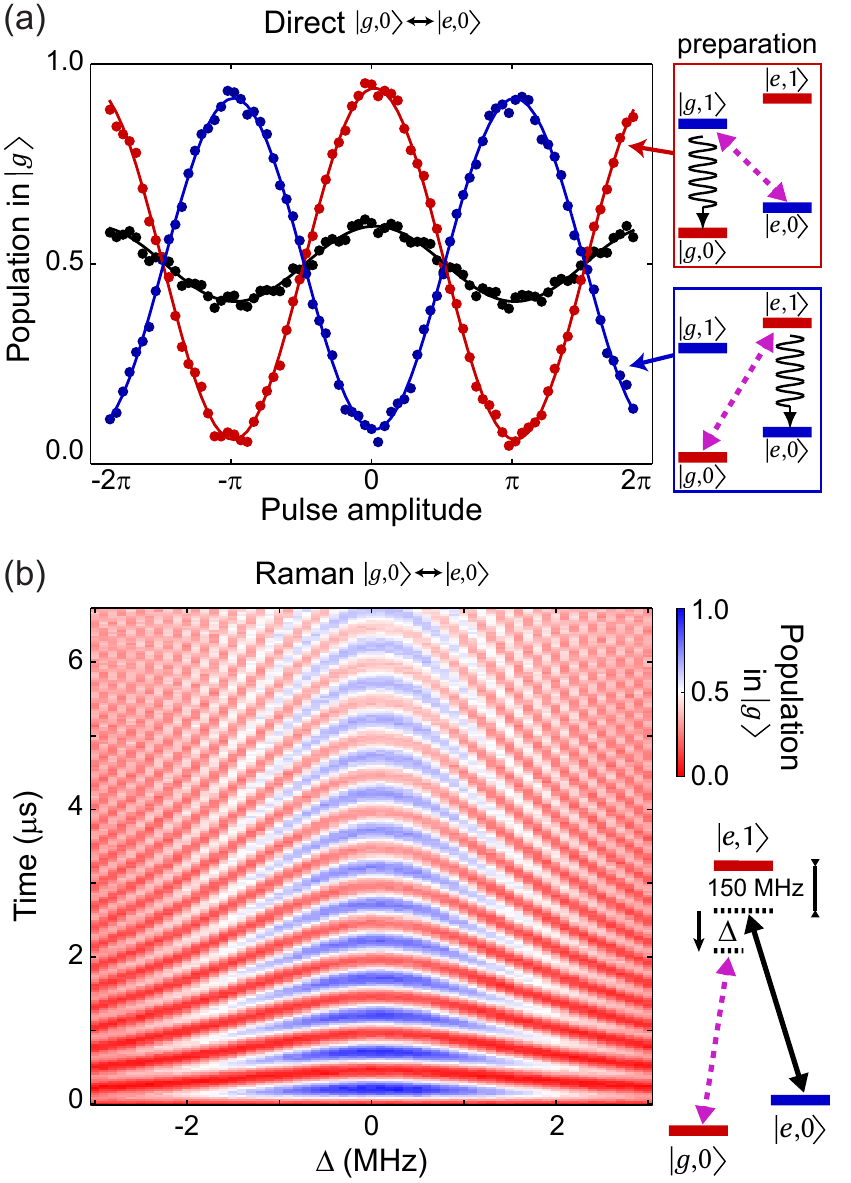}
\caption{\label{fig4}  (a) Rabi flops of the fluxonium $\ket{g,0}\leftrightarrow\ket{e,0}$ transition with different initial preparations before applying a 20-ns $\sigma$ pulse at $f_q$. The black dots correspond to the qubit starting in thermal equilibrium, where we measured the qubit to be $60\%$ in $\ket{g}$, corresponding to $62~\mathrm{mK}$. The red dots correspond to initially cooling the qubit to $\ket{g}$ by applying a tone resonant with the $\ket{e,0}\leftrightarrow \ket{g,1}$ transition (see inset). The qubit then is in $\ket{g}$ with a $94\%$ probability. The blue dots correspond to initially preparing the qubit in $\ket{e}$ by applying a tone resonant with the $\ket{g,0}\leftrightarrow \ket{e,1}$ transition (see inset), which results in a $91.5\%$ probability for the qubit to be in $\ket{e}$. The solid lines are sinusoidal fits to the measured Rabi oscillations. (b) Rabi oscillations of the fluxonium $\ket{g,0}\leftrightarrow\ket{e,0}$ transition by a Raman process through the $\ket{e,1}$ state (see inset). The $\ket{e,0}\leftrightarrow \ket{e,1}$ tone is applied $150~\mathrm{MHz}$ detuned from resonance, and the $\ket{g,0}\leftrightarrow\ket{e,1}$ is applied at $150~\mathrm{MHz}+ \Delta$.}
\end{figure}

We demonstrated that our circuit can be treated as a $\Lambda$-type system by performing both incoherent and coherent operations using Raman transitions. All measurements were performed at the $\Phi^f_\mathrm{ext}/\Phi_0 = 6.5$ fluxonium sweet spot. We first demonstrate cooling of the fluxonium artificial atom by spontaneous Raman scattering. The black dots in Fig.~\ref{fig4}a correspond to a standard amplitude-Rabi experiment on the fluxonium qubit, where a Gaussian pulse of 20-ns $\sigma$ width and varying amplitude was applied at $f_q$. As the amplitude was varied, the qubit population oscillated between thermal equilibrium and inverted population. From the oscillation amplitude we inferred that the qubit had $60\%$ probability to be in the ground state $\ket{g}$, which corresponds to a temperature of $62~\mathrm{mK}$. This experiment was repeated after applying a tone resonant with the $\ket{e,0}\leftrightarrow\ket{g,1}$ transition for $5~\mathrm{\mu s}$. After this duration, the fluxonium had $94\%$ probability to be in $\ket{g}$, which corresponds to $9~\mathrm{mK}$ - well below thermal equilibrium. We have thus demonstrated cooling of the fluxonium to its ground state by the Raman process shown in the inset of Fig.~\ref{fig4}a. We also initialized the qubit in $\ket{e}$ by applying a tone resonant with the $\ket{g,0}\leftrightarrow\ket{e,1}$ transition before performing the Rabi experiment (blue). This inverted the fluxonium population and prepared it in $\ket{e}$ with $91.5\%$ probability. The ground state population was calibrated from qubit measurements in conjunction with a theoretical model for Raman cooling (see Appendix D). 

We were also able to coherently control the qubit through the $\Lambda$-type system. Figure~\ref{fig4}b shows Rabi oscillations of the fluxonium qubit via a virtual transition through the $\ket{e,1}$ state. Two tones were applied at the qubit for varying lengths of time after initially preparing the fluxonium ground state using Raman cooling. One was detuned $150~\mathrm{MHz}$ below the resonator frequency (black arrow in the inset of Fig.~\ref{fig4}b), and the other was detuned below the $\ket{g,0}\leftrightarrow\ket{e,1}$ transition by $(150 + \Delta)~\mathrm{MHz}$ where $\Delta$ was a variable additional detuning (magenta arrow in the inset of Fig.~\ref{fig4}b). We observed a typical Rabi oscillation pattern, which shows that we can control a qubit with transition frequency $f_q = 500~\mathrm{MHz}$ by only applying tones around $7~\mathrm{GHz}$. Note that the optimal detuning of the $\ket{g,0}\leftrightarrow\ket{e,1}$ transition corresponds to $\Delta = 60~\mathrm{kHz}$, due to a Stark shift of this transition. From $\Delta$, we can extract a drive amplitude of $g_3 /2 \pi = 3~\mathrm{MHz}$ for the $\ket{g,0}\leftrightarrow\ket{e,1}$ transition due to nonlinear coupling (see Appendix E).

The methods described above extend the quantum control of atomic physics by Raman transitions to superconducting circuits. In atomic physics, these transitions couple levels whose direct transition is forbidden and thus protected from environmental noise. However, the circumstances in which this idea can be exploited only occur in a limited number of atoms. By contrast, in superconducting artificial atoms we can engineer circuits to implement the transitions needed for this quantum control. Here, as a proof-of-principle, we applied this method to the fluxonium artificial atom, whose $\ket{g}\leftrightarrow \ket{e}$ transition can be directly driven. This Raman control is absolutely essential for more complex physically-protected qubits~\cite{doucot_physical_2012,brooks_protected_2013,bell_protected_2014,kou_fluxonium-based_2017}, however, whose inherent protection makes them impossible to directly control and read out. Note also that our technique separates qubit control from the qubit transition frequency. Previous two-photon implementations~\cite{wallraff_sideband_2007,novikov_raman_2016} of the $\ket{g,0} \leftrightarrow \ket{e,1}$ transition relied on the direct $\ket{g,0} \leftrightarrow \ket{e,0}$ transition, and thus cannot be used for the control of protected qubits. With our method, one can perform coherent operations on the qubit while its direct transition remains completely isolated.

\section{V. Conclusions}
In conclusion, nonlinear coupling between a qubit and a resonator can be used to directly drive transitions forbidden by parity symmetry at the fluxonium sweet spot. This method implements a $\Lambda$-type system within superconducting circuits. We created a new transition which completes the triangle spanning the $\ket{g,0}$, $\ket{e,0}$, and $\ket{e,1}$ states of the fluxonium qubit coupled to an antenna resonator. We then demonstrated cooling of the fluxonium qubit by spontaneous Raman scattering, and coherent oscillations between the ground and excited state driven by a stimulated Raman process. 

Furthermore, the ability to create new transitions in a superconducting circuit opens the door to applications in microwave quantum optics~\cite{you_atomic_2011} and autonomous error correction~\cite{kerckhoff_designing_2010}. While we have focused here on $\Lambda$-system physics, our ability to drive a two-excitation transition can also be understood as tunable mode coupling, useful for the single-drive implementation of protocols for photon detection~\cite{inomata_single_2016} and remote entanglement~\cite{campagne-ibarcq_deterministic_2017,axline_-demand_2017,kurpiers_deterministic_2017}.

\section{Acknowledgments}
We acknowledge fruitful discussions with Liang Jiang.
Facilities use was supported by YINQE, the Yale SEAS cleanroom, and NSF MRSEC DMR 1119826. This research was supported by ARO under Grant No. W911NF-14-1-0011, by MURI-ONR Grant No. N00014-16-1-2270, and by NSF DMR-1609326. 

\section{Appendices}
\subsection{A. Theoretical description of a SNAIL}

A sketch of a SNAIL circuit element can be seen in Fig.~2b. A detailed derivation and analysis of this circuit is given in Refs.~\onlinecite{frattini_3-wave_2017,vool_engineering_2017}, and in this Appendix we only give a brief overview.
We can express the potential (inductive) energy of the SNAIL as:
\begin{equation}
\begin{split}
\bm{U_\mathrm{SNAIL}}(\bm{\varphi}) = &-\alpha E_\mathrm{J} \cos \bm{\varphi} \\
&-n E_\mathrm{J} \cos \left( \frac{\Phi^\mathrm{S}_\mathrm{ext}/\phi_0-\bm{\varphi}} {n}\right),
\end{split}
\label{eq:SNAIL_pot}
\end{equation}
where $\bm{\varphi}$ is the superconducting phase across the small junction of the SNAIL, $E_\mathrm{J}$ is the Josephson energy of the large SNAIL junction, $\alpha$ is the ratio between the small and large junction, $n$ is the number of large junctions in the SNAIL loop (in our implementation $n=3$), $\Phi^\mathrm{S}_\mathrm{ext}$ is the external flux through the SNAIL loop and $\phi_0$ is the reduced magnetic flux quantum. In this description we have eliminated the dynamics of the modes within the $n$-junction array, and consider the circuit as a single degree of freedom with equal phases across the array junctions. 

To expand Eq.~\ref{eq:SNAIL_pot} as a non-linear inductor, we must first find the minimum $\varphi_\mathrm{min}$ of the potential. This minimum depends on $\Phi^\mathrm{S}_\mathrm{ext}$ and $\alpha$, and can be numerically obtained for each of those. Then we can expand the SNAIL potential around the minimum, using the new coordinate $\bm{\widetilde \varphi} = \bm\varphi - \varphi_\mathrm{min}$. We express the Taylor expansion of the potential as:
\begin{equation}
\bm{U_\mathrm{SNAIL}}(\bm{\widetilde\varphi}) = c_2 \bm{\widetilde \varphi}^2 + c_3 \bm{\widetilde \varphi}^3 +c_4 \bm{\widetilde \varphi}^4 +...,
\label{eq:SNAIL_expand}
\end{equation}
where $c_m$ is the coefficient of the $m$-th order in the expansion. $c_2$ is related to the linear inductance of the SNAIL as $L_\mathrm{S} = \frac{\phi_0^2}{2 c_2 E_\mathrm{J}}$. 
These coefficients also depend on $\Phi^\mathrm{S}_\mathrm{ext}$, $\alpha$, and $n$, and can be obtained numerically. 

\subsection{B. Theoretical description of the device}

\begin{figure*} [htp]
\centering
\centerline{\includegraphics[angle = 0]{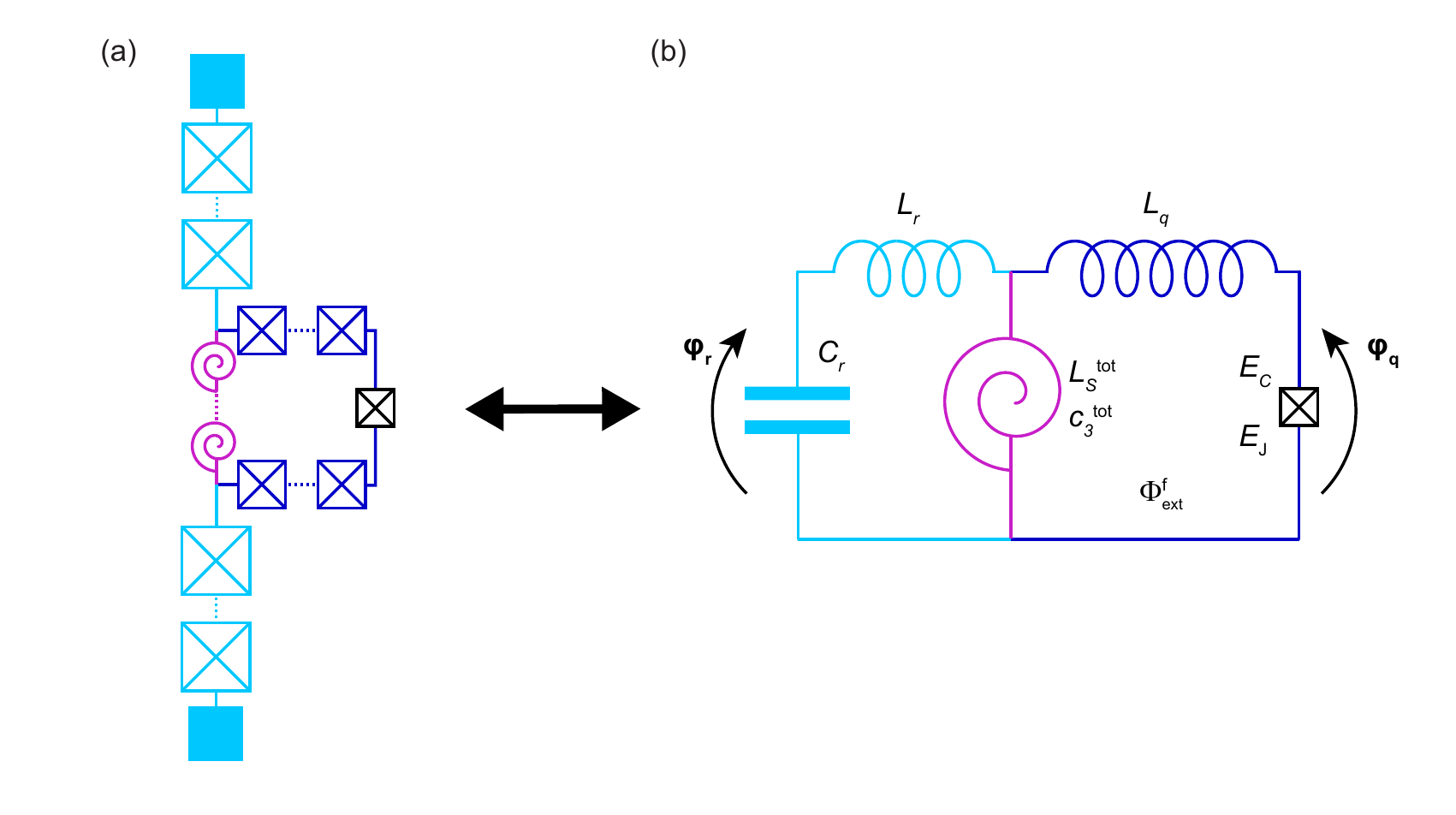}}
\caption{(a) The fluxonium artificial atom  is made up of a small junction (black) shunted by a large inductance, itself made of an array of larger Josephson junctions (dark blue). Some of this inductance is shared with the resonator (light blue). In this design, the shared elements (magenta) are taken to be SNAILs, which allows us to break selection rules at the fluxonium sweet spot. (b) The equivalent circuit diagram. This circuit has two DOF, and we express the Hamiltonian as a function of $\bm{\varphi_r}$ - the flux across the resonator capacitor and $\bm{\varphi_q}$, the flux across the fluxonium small junction.}
\label{fig:snail_fluxonium_circuit}
\end{figure*}

Fig.~2a shows a sketch of our circuit in which a fluxonium artificial atom is coupled to an antenna resonator by sharing a nonlinear inductance composed of SNAILs. To understand the behavior of the circuit quantitatively, let us simplify it into an effective circuit given in Fig.~\ref{fig:snail_fluxonium_circuit}b. The fluxonium is now represented by a small junction (black) with Josephson energy $E_\mathrm{J}$ and capacitive energy $E_C$, shunted by a linear inductance $L_q$ (dark blue). An external flux $\Phi^{f}_\mathrm{ext}$ is threaded through the fluxonium loop. The resonator (light blue) is represented as an LC oscillator with capacitance $C_r$ and unshared inductance $L_r$. The $N$ SNAIL array which couples the two systems (in our implementation $N=5$) is reduced to a single effective SNAIL (magenta). We represent the SNAIL as having only second-order and third-order terms.

We can label the superconducting phase across the SNAIL array as $\bm{\varphi_\mathrm{S}}$, and assume it is divided equally across all $N$ SNAILs in the array. Thus, we can calculate the coefficients of the total SNAIL array from those of the SNAIL. 
\begin{equation}
c^\mathrm{tot}_2 \bm{\varphi_\mathrm{S}}^2= N  c_2  \left(\frac{\bm{\varphi_\mathrm{S}}}{N}\right)^2, 
\end{equation}
where $c_2$ is the second-order coefficient of a single SNAIL in the array, and $c^\mathrm{tot}_2$ is the second-order coefficient of the whole array. Thus, $c^\mathrm{tot}_2 = c_2/N$ or $L^\mathrm{tot}_\mathrm{S} = N L_\mathrm{S}$. Linear inductances in series are simply added, as expected. A similar calculation shows that $c^\mathrm{tot}_3 = \frac{c_3}{N^2}$, and so the third-order non-linearity is suppressed by an additional factor of $N$. Higher order non-linearities are similarly suppressed by higher and higher factors, making the low-order non-linearity assumption better.

The circuit in Fig.~\ref{fig:snail_fluxonium_circuit}b has only two true degrees of freedom and we choose to use the phase across the Josephson junction $\bm{\varphi_q}$, and the phase across the resonator capacitance $\bm{\varphi_r}$. These are related to the fluxonium and resonator flux operators given by $\bm{\varphi_q}=\bm{\phi_q}/\phi_0$ and  $\bm{\varphi_r}=\bm{\phi_r}/\phi_0$, where $\phi_0$ is the reduced magnetic flux quantum.

We can derive the Hamiltonian for the circuit by following the circuit quantization protocol~\cite{vool_introduction_2017}. A very similar derivation is given in Ref.~\onlinecite{smith_quantization_2016}, with a shared linear inductance replacing the SNAIL. The addition of the SNAIL adds a three-wave mixing term to the simple fluxonium-resonator Hamiltonian, of the form:
\begin{equation}
\bm{H_\mathrm{3WM}} = c_3^\mathrm{tot} \left(\frac{L_r L^\mathrm{tot}_\mathrm{S}}{L_q (L_r+L^\mathrm{tot}_\mathrm{S})} \bm{\varphi_q} + \frac{L^\mathrm{tot}_\mathrm{S}}{L_r + L^\mathrm{tot}_\mathrm{S}} \bm{\varphi_r}\right)^3,
\end{equation}
where we have assumed that $L_q \gg L_r, L^\mathrm{tot}_\mathrm{S}$. This three-wave mixing Hamiltonian gives rise to several effects through its different mixing terms, but let us focus on two terms of special importance:
\begin{eqnarray}
\bm{H_\mathrm{\ket{g}-\ket{f}}} &=& 3 c_3^\mathrm{tot}  \left(\frac{L^\mathrm{tot}_\mathrm{S}}{L_q} \right)^2 \frac{L_r^2 L^\mathrm{tot}_\mathrm{S}}{(L_r + L^\mathrm{tot}_\mathrm{S})^3} \bm{\varphi_r}\bm{\varphi_q}^2 \label{eq:g_f_Ham}\\
\bm{H_\mathrm{\ket{g,0}-\ket{e,1}}} &=& 3 c_3^\mathrm{tot}  \frac{L^\mathrm{tot}_\mathrm{S}}{L_q}  \frac{L_r (L^\mathrm{tot}_\mathrm{S})^2}{(L_r + L^\mathrm{tot}_\mathrm{S})^3} \bm{\varphi_r}^2\bm{\varphi_q} \label{eq:g0_e1_Ham}
\end{eqnarray}

The term in Eq.~\ref{eq:g_f_Ham} is proportional to the term $\bm{\varphi_r}\bm{\varphi_q}^2$. With an additional resonator drive, this gives rise to an even drive term of the form $\bm{\varphi_q}^2$ which is able to drive the fluxonium $\ket{g}\leftrightarrow\ket{f}$ transition at the fluxonium sweet spot as $\bra{g}\bm{\varphi_q}^2\ket{f} \neq 0$. 
To understand this effective coupling term, first notice that from circuit quantization $\bm{\varphi_r} = \varphi^r_\mathrm{ZPF}(\bm{a_r}+\bm{a^\dagger_r})$ where $a_r$ is the resonator decay operator and $\varphi^r_\mathrm{ZPF}$ are the zero-point fluctuations of the resonator phase operator. We can now add a drive term of the form $\epsilon (\bm{a_r}+\bm{a^\dagger_r})$ in the drive frequency rotating frame, where $\epsilon$ is the resonator drive amplitude. By applying the displacement operator, we end up with a drive term of the form $\varphi^r_\mathrm{ZPF} \alpha_r \bm{\varphi_q}^2$ where $\alpha_r$ is the coherent state amplitude in the resonator. 

 The term in Eq.~\ref{eq:g0_e1_Ham} similarly leads to a drive term of the form $\bm{\varphi_q}\bm{\varphi_r}$. This is another even term, but one that allows us to drive the forbidden joint transitions such as $\ket{g,0}\leftrightarrow\ket{e,1}$ and $\ket{e,0}\leftrightarrow\ket{g,1}$\footnote{The distinction between $\bm{\varphi_q}^2$ and $\bm{\varphi_q}\bm{\varphi_r}$ and the terms they can drive is not absolute, due to the coupling between the resonator and the fluxonium. Both of these drive terms can drive the transitions $\ket{g}\leftrightarrow\ket{f}$ and $\ket{g,0}\leftrightarrow\ket{e,1}$ to some extent. However, as the coupling between the resonator and the fluxonium is weak (we are in the dispersive regime), we associate each term with the transitions it couples to more strongly.}. 
 Eq.~\ref{eq:g0_e1_Ham} should thus remind us of tunable mode coupling, as it gives rise to both a beam-splitter term which enables the $\ket{e,0}\leftrightarrow \ket{g,1}$ transition, and a two-mode squeezing term which enables the $\ket{g,0} \leftrightarrow \ket{e,1}$ transition. Thus, our selection-rule-breaking drive can also be understood as a tunable coupling between modes, such that the parity is preserved.

There are two important things to notice in the coefficients of Eqs.~\ref{eq:g_f_Ham} and \ref{eq:g0_e1_Ham}. First of all, they both depend on $L_r$ in the numerator. $L_r$ is the unshared resonator inductance, and thus one would expect that if the antenna shares more of its inductance, it is more coupled to the fluxonium and thus the SNAIL is better able to drive the forbidden transitions. This intuition is false, as when $L_r=0$, the SNAIL element is the entire inductance of the resonator and the phase across it is $\bm{\varphi_r}$. There is still coupling between the two modes, mediated by the $L_q$ inductor, but the SNAIL does not participate in it and thus there is no three-wave mixing for the qubit mode. Thus, a substantial $L_r$, comparable to $L^\mathrm{tot}_\mathrm{S}$, is necessary to drive forbidden transitions.

The second thing to notice is that the coefficient in Eq.~\ref{eq:g0_e1_Ham} is larger than Eq.~\ref{eq:g_f_Ham} by a factor of $L_q/L^\mathrm{tot}_\mathrm{S}$, which is experimentally $\approx 50$. Thus, this coupling scheme is more suited to drive the two-mode forbidden transitions such as $\ket{g,0} \leftrightarrow \ket{e,1}$ and $\ket{e,0} \leftrightarrow \ket{g,1}$.

\subsection{C. Calculating the drive amplitude $g_3$}

The term in Eq.~\ref{eq:g0_e1_Ham} allows us to quantify the effective coupling strength of our nonlinear transitions. Let us take this term, and add a direct drive on the resonator of the form $\epsilon (\bm{a_r}+\bm{a^\dagger_r})$. As mentioned in Appendix B, we can displace the resonator by the transformation $\bm{a_r} \rightarrow \bm{a_r} + \alpha_r$, where the displacement is chosen to be the coherent steady-state amplitude in the resonator $\alpha_r = \frac{\epsilon}{i \kappa/2- \Delta_r}$ where $\Delta_r$ is the drive detuning from the resonator resonance frequency and $\kappa$ is its linewidth. This transformation eliminates the direct drive term and we end up with an effectively undriven resonator. 

The term in Eq.~\ref{eq:g0_e1_Ham}, however, in this new frame gives rise to the effective coupling term that we require. This effective term is of the form:
\begin{equation}
\bm{H_\mathrm{eff}} = 6 \varphi^r_\mathrm{ZPF} \alpha_r c_3^\mathrm{tot} \phi_0^3 \frac{L^\mathrm{tot}_\mathrm{S}}{L_q}  \frac{L_r (L^\mathrm{tot}_\mathrm{S})^2}{(L_r + L^\mathrm{tot}_\mathrm{S})^3} \bm{\varphi_r}\bm{\varphi_q},
\label{eq:g3effham}
\end{equation}
and drive amplitude term can then be simply obtained by $g_3 = \bra{g,0}\bm{H_\mathrm{eff}}\ket{e,1}$.

\begin{figure} [htp]
\centering
\centerline{\includegraphics[angle = 0]{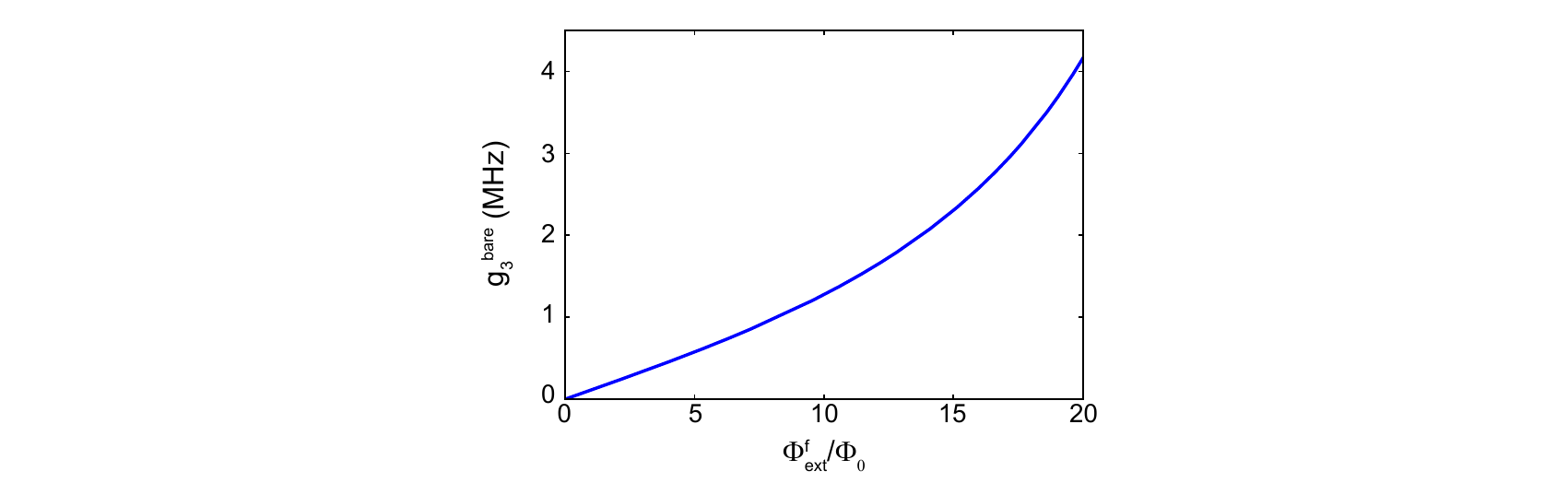}}
\caption{The bare coupling coefficient $g^\mathrm{bare}_3$ (see Eq.~\ref{eq:g3bare}) vs. the external flux through the fluxonium loop $\Phi^f_\mathrm{ext}$.}
\label{fig:supfig2}
\end{figure}

We can separate the discussion of this term into three components. The first one is the bare coupling term:
\begin{equation}
g^\mathrm{bare}_3 = 6 \varphi^r_\mathrm{ZPF} c_3^\mathrm{tot} \frac{L^\mathrm{tot}_\mathrm{S}}{L_q}  \frac{L_r (L^\mathrm{tot}_\mathrm{S})^2}{(L_r + L^\mathrm{tot}_\mathrm{S})^3},
\label{eq:g3bare}
\end{equation}
which includes the specific parameters of the design which we have discussed in Appendix B. The flux $\Phi^\mathrm{S}_\mathrm{ext}$ through the SNAIL strongly influences the value of $c_3^\mathrm{tot}$, and to a lesser extent $L^\mathrm{tot}_\mathrm{S}$. Thus $g^\mathrm{bare}_3$ is responsible for the improved ability to drive the forbidden transition with increased flux (see Fig.~3). Fig.~\ref{fig:supfig2} shows $g^\mathrm{bare}_3$ as a function of the flux through the fluxonium loop ,$\Phi^f_\mathrm{ext}$. Recall that due to the difference in loop areas, $\Phi^\mathrm{S}_\mathrm{ext} = \Phi^f_\mathrm{ext} / 60$. 

The second component in the $g_3$ term is the matrix element $\bra{g,0}\bm{\varphi_r}\bm{\varphi_q}\ket{e,1}$. Note that $\bm{\varphi_r}$ and $\bm{\varphi_q}$ are \textbf{not} the field operators of the resonator and fluxonium modes, but simply convenient bases made up of their linear parts. Especially, $\bm{\varphi_q}$ is the field operator of a linear mode very different from the fluxonium qubit. The statements $\bra{g}\bm{\varphi_q}\ket{f} = 0$ and $\bra{g}\bm{\varphi_q}^2\ket{f} \neq 0$ are true due to the selection rules, but calculating the value of the matrix element requires a diagonalization of the fluxonium Hamiltonian, and is usually done numerically~\cite{smith_quantization_2016}. However, this matrix element is identical for all fluxonium sweet spots $\Phi^f_\mathrm{ext}/\Phi_0 = m + 1/2$ where $m \in \mathbb{N}$. Thus, is can be treated as a constant for the purposes of this paper and from a numerical diagonalization of the Hamiltonian we obtain $\bra{g,0}\bm{\varphi_r}\bm{\varphi_q}\ket{e,1} \simeq 2$ at the fluxonium $\Phi_0/2$ sweet spots. It is also important to note that $\bra{g,0}\bm{\varphi_r}\bm{\varphi_q}\ket{e,1} = \bra{e,0}\bm{\varphi_r}\bm{\varphi_q}\ket{g,1}$ and thus our drive term can excite $\ket{e,0}\leftrightarrow \ket{g,1}$ and $\ket{g,0}\leftrightarrow \ket{e,1}$ with equal amplitude, and the selection is made by the transition frequency to which we tune our external drive.

The last component is the $\alpha_r= \frac{\epsilon}{i \kappa/2- \Delta_r}$, the coherent population in the cavity during the drive. This term is proportional to $\epsilon$ and thus shows the increase in transition rate as our drive amplitude increases. This is the component we cannot estimate from system parameters, thus limiting our ability to predict the transition rate of the nonlinear transition. However, we have independent calibrations for the value of $g_3$ from the spontaneous and stimulated Raman transition measurements, as will be discussed in Appendix D, and we can use them to estimate the population of the resonator.

Note, however, that we can directly calculate the ratio $g_3/\epsilon$, which is the ratio of the rate at which we drive forbidden transitions such as $\ket{g,0}\leftrightarrow \ket{e,1}$ and $\ket{e,0}\leftrightarrow \ket{g,1}$, and the rate at which we directly drive the resonator transitions such as $\ket{g,0}\leftrightarrow \ket{g,1}$ or $\ket{e,0}\leftrightarrow \ket{e,1}$. From our system parameters at $\Phi^f_\mathrm{ext} = 6.5 \Phi_0$, we can estimate $g_3/\epsilon = 0.003$. This rate comparison assumes the transitions are driven using a tone of the same drive amplitude, at a frequency equal to the resonance frequency for each transition.

\subsection{D. Calibration of the fluxonium ground state population}

In Fig.~4 we present the measurements of spontaneous and stimulated Raman transitions in terms of the fluxonium ground state population. This axis is actually calibrated by using known parameters and by assuming a Raman cooling model for the results in Fig.~4a, and we explain this calibration in detail in this Appendix.

Recall that we measure the state of the fluxonium via its effect on the resonator frequency, and there are positions in the $I$-$Q$ phase-space of the reflected signal which correspond to the fluxonium being in $\ket{g}$ and $\ket{e}$. Let us mark half of the distance in phase-space between these two positions as $A$. Thus, if the initial state of the fluxonium qubit is exactly $\ket{g}$ and it performs perfect Rabi oscillations, the amplitude for these observed oscillations would be $A$. However, as our qubit is in thermal equilibrium, the actual measured amplitude is: 
\begin{equation}
A_\mathrm{th} = A(P^g_\mathrm{th}-P^e_\mathrm{th}) = A(2P^g_\mathrm{th}-1) \label{eq:amp_th},
\end{equation}
where $P^g_\mathrm{th}$ ($P^e_\mathrm{th}$) is the probability the fluxonium is $\ket{g}$ ($\ket{e}$) in thermal equilibrium. Note that $A_\mathrm{th}$ is the amplitude of oscillations measured in the black curve in Fig.~4a.

Similarly, we can define the probability in $\ket{g}$ after the $\ket{e,0}\leftrightarrow \ket{g,1}$ (red) Raman cooling sequence as $P^g_\mathrm{red}$, and the probability in $\ket{e}$ after the $\ket{g,0}\leftrightarrow \ket{e,1}$ (blue) Raman cooling sequence as $P^e_\mathrm{blue}$. Their corresponding Rabi oscillation amplitudes are then:
\begin{eqnarray}
A_\mathrm{red} &=& A(2P^g_\mathrm{red}-1) \label{eq:amp_red}\\
A_\mathrm{blue} &=& A(2P^e_\mathrm{blue}-1) \label{eq:amp_blue}
\end{eqnarray}

We can also find expressions for $P^g_\mathrm{red}$ and $P^e_\mathrm{blue}$. Let us label the transition rate of the $\ket{e,0}\leftrightarrow \ket{g,1}$ transition as $g_\mathrm{red}$.  The Raman cooling thus involves a coherent excitation to the state $\ket{g,1}$  with a rate $g_\mathrm{red}$, followed by an incoherent decay of the resonator to $\ket{g,0}$ at rate $\kappa$. As our resonator has a large decay rate $\kappa = 2\pi\times 16.8~\mathrm{MHz}$, we can reasonably assume that $g_\mathrm{red} \ll \kappa$. We can thus adiabatically eliminate the higher state. A similar process can be done for the $\ket{g,0}\leftrightarrow \ket{e,1}$ transition and its rate $g_\mathrm{blue}$. Note that we can assume $g_\mathrm{red} = g_\mathrm{blue} = g_3$ as their corresponding matrix elements are identical, as discussed in Appendix D. Thus, we express the cooling rate for both processes using adiabatic elimination:
\begin{equation}
\Gamma_\mathrm{cool} = \frac{4g_3^2}{\kappa}.
\end{equation}

The thermal fluxonium population can be described in terms of an ``up" rate $\Gamma_\uparrow$ which is the rate of transition $\ket{g}\rightarrow\ket{e}$, and a ``down" rate $\Gamma_\downarrow$ which is the rate of transition $\ket{e}\rightarrow\ket{g}$. Their sum equals the total thermalization rate $\Gamma_\downarrow + \Gamma_\uparrow = \Gamma_1$, and the qubit population is related to them via:
\begin{equation}
P^g_\mathrm{th} = \frac{\Gamma_\downarrow}{\Gamma_\downarrow+\Gamma_\uparrow} = \frac{\Gamma_\downarrow}{\Gamma_1}
\end{equation}
from a detailed balance assumption in equilibrium.

The Raman cooling tones then enter to aid the different thermal equilibration rates. The red tone cools the qubit to $\ket{g,0}$, and thus the cooling rate $\Gamma_\mathrm{cool}$ aids $\Gamma_\downarrow$. Similarly, the blue tone cools the qubit to $\ket{e,0}$, and thus the cooling rate $\Gamma_\mathrm{cool}$ aids $\Gamma_\uparrow$. We can thus express the populations after cooling as:
\begin{eqnarray}
P^g_\mathrm{red} &=&\frac{\Gamma_\mathrm{cool} + \Gamma_\downarrow}{\Gamma_\mathrm{cool} + \Gamma_\downarrow + \Gamma_\uparrow} = \frac{4g_3^2 + \kappa\Gamma_\downarrow}{4g_3^2 + \kappa\Gamma_\downarrow + \kappa\Gamma_\uparrow}\\
P^e_\mathrm{blue} &=& \frac{\Gamma_\mathrm{cool} + \Gamma_\uparrow}{\Gamma_\mathrm{cool} + \Gamma_\downarrow + \Gamma_\uparrow} = \frac{4g_3^2 + \kappa\Gamma_\uparrow}{4g_3^2 + \kappa\Gamma_\downarrow + \kappa\Gamma_\uparrow}
\end{eqnarray}

Let us summarize all these relations. Eqs.~\ref{eq:amp_th}, \ref{eq:amp_red}, and \ref{eq:amp_blue} relate three measured quantities, $A_\mathrm{th}$, $A_\mathrm{red}$, and $A_\mathrm{blue}$, to expressions with several unknowns. From the following equations, we see that we have expressed all of these terms using only three unknowns: $A$, $g_3$, and $P^g_\mathrm{th}$. All other unknowns can be expressed using these three, as well as known quantities such as $\kappa = 2\pi\times 16.8~\mathrm{MHz}$ and $\Gamma_1$ which is related to the qubit lifetime $1/\Gamma_1 = T_1 = 5.7~\mathrm{\mu s}$. Thus, we can solve a set of three equations with three unknowns, and extract the thermal population of our fluxonium qubit.

The extracted qubit equilibrium temperature is $62~\mathrm{mK}$, which corresponds to $P^g_\mathrm{th} = 0.6$. We also obtain the population after cooling to $\ket{g}$, $P^g_\mathrm{red} = 0.94$, and the population after cooling to $\ket{e}$, $P^e_\mathrm{blue} = 0.915$. This is the calibration of the qubit population which is used in Fig.~4.

This analysis also gives us the transition rate, $g_3 = 2\pi\times0.87~\mathrm{MHz}$. Notice we self-consistently justify our assumption $g_3 \ll \kappa$. We can also compare this measurement to the theoretical prediction. In Appendix D we discussed the calculation of the rate $g_3$ (see Eq.~\ref{eq:g3effham}), where we can independently predict all the coefficients besides $\alpha_r$, which is the coherent state population in the resonator which enables this drive. From the measurement of $g_3$, we can estimate the photon population in the resonator due to this cooling drive as $|\alpha_r|^2 = 0.35$.

\subsection{E. Estimation of drive rates from stimulated Raman transition measurements}

Notice that the Rabi-oscillations in Fig.~4b are not quite centered at $\Delta = 0$, but rather are slightly offset at $\Delta = 60 ~\mathrm{kHz}$. This is a result of the Stark shift in the nonlinear mode, and it is related to the drive amplitude and the detuning by $\Delta_\mathrm{Stark} = \frac{g_3^2}{\Delta_r}$. Notice that in this case there is only one nonlinear mode, and only it experiences a Stark shift.

This allows us to estimate $g_3 = 2\pi\times 3~\mathrm{MHz}$. Similarly to the procedure at the end of Appendix D, we can thus estimate the photon number in the resonator to be $|\alpha_r|^2 = 4.3$. This seems to contradict the transition $\ket{g,0}\leftrightarrow \ket{e,1}$ as there are more photons in the cavity. But note that while this drive is on we are in a displaced frame, and the states $\ket{g,0}$ and $\ket{e,1}$ are defined from this displaced value. Also note this value is larger than that of the cooling drive by a factor of $\approx 10$, consistent with the generator being set $10 ~\mathrm{dB}$ higher for this measurement. This is also the basis of the rate estimate in Fig.~2, accounting for the change in the external drive amplitude for the spectroscopy measurement.

\begin{figure} [htp]
\centering
\centerline{\includegraphics[angle = 0]{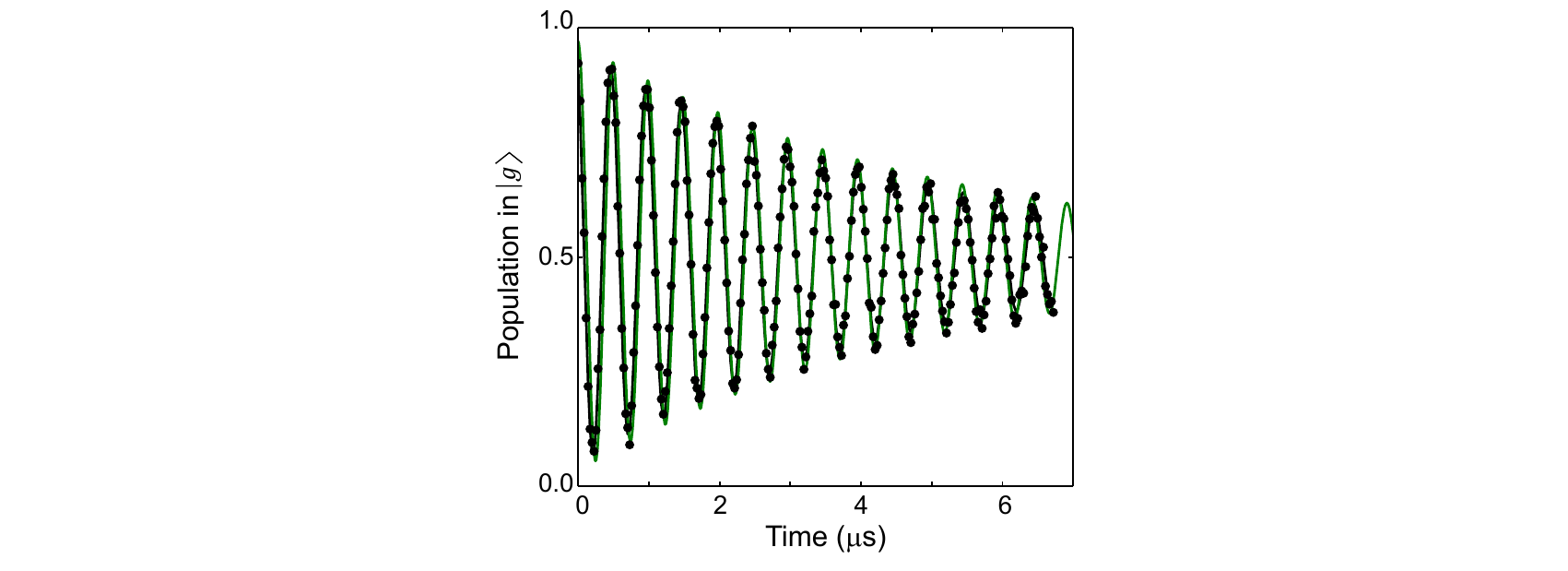}}
\caption{Rabi oscillations of the fluxonium qubit due to stimulated Raman transitions. This is a cut of Fig.~4b at $\Delta = 100~\mathrm{kHz}$. The green line shows a fit to Eq.~\ref{eq:lambda_ham} with only one fit parameter $\epsilon$}
\label{fig:supfig3}
\end{figure}

Fig.~\ref{fig:supfig3} shows a cut of Fig.~4b at $\Delta = 100~\mathrm{kHz}$. To quantify these oscillations, we can compare then to a theoretical model given by:
\begin{equation}
\begin{split}
\bm{H_\Lambda}/\hbar &= \omega_r \bm{a^\dagger_r a_r} + \frac{\omega_q}{2}\bm{\sigma_z} + \frac{\chi}{2}\bm{a^\dagger_r a_r}\bm{\sigma_z} \\
&+ 2\epsilon \cos (\omega_d t) (\bm{a_r}+ \bm{a^\dagger_r}) \\
&+ 2g_3 \cos (\omega_{nl} t) (\bm{a_r\sigma_-}+\bm{a^\dagger_r \sigma_+}),
\end{split}
\end{equation}
where $\bm{a_r}$ is the resonator annihilation operator and the fluxonium is modeled as a two-level system in the Pauli $\bm{\sigma_z}$ basis. $g_3$ is the transition rate of the nonlinear drive (same coefficient as discussed in Appendix C), $\epsilon$ is the coefficient of the direct resonator drive and $\chi = 2\pi\times 0.7~\mathrm{MHz}$ is the dispersive coupling between the fluxonium and the resonator (estimate from the $I$-$Q$ response of the fluxonium). $\omega_d = \omega_r - \Delta_r$ is the drive frequency of the direct cavity drive where $\omega_r$ is the resonator frequency and $\Delta_r = 2\pi\times 150~\mathrm{MHz}$ is the drive detuning. $\omega_{nl} = \omega_r+\omega_q - \Delta_r - \Delta$ is the drive frequency for the nonlinear transition, and it is detuned from the $\ket{g,0}\leftrightarrow \ket{e,1}$ resonance frequency by $\Delta_r + \Delta$ where $\Delta$ is a variable detuning (see Fig.~4b).

By moving to the rotating frames  $\bm{U_r}=e^{i\bm{a^{\dagger}_r a_r}\omega_{d}t}$, $\bm{U_q}=e^{i\bm{\sigma_{z}}\frac{\omega_{q}-\Delta}{2}t}$ and taking the rotating wave approximation, we arrive at the time-independent Hamiltonian:
\begin{equation}
\begin{split}
\bm{H_\Lambda}/\hbar &= \Delta_r \bm{a^\dagger_r a_r} + \frac{\Delta}{2}\bm{\sigma_z} + \frac{\chi}{2}\bm{a^\dagger_r a_r}\bm{\sigma_z} \\
&+ \epsilon(\bm{a_r}+\bm{a^\dagger_r}) + g_3(\bm{a_r\sigma_-}+\bm{a^\dagger_r \sigma_+}).
\end{split}
\label{eq:lambda_ham}
\end{equation}

Notice that we have independent measurements of every coefficient in Eq.~\ref{eq:lambda_ham} except $\epsilon$. We also know all the decay constants for the fluxonium and resonator, and the initial population of the fluxonium (which was cooled to $94\%$ in $\ket{g,0}$ by Raman cooling). Thus, we can simulate the master equation for our system, and fit it to our measurement in Fig.~\ref{fig:supfig3} with only a single fit parameter $\epsilon$. This numerical simulation result is shown as the green line in the figure. Notice we obtain good agreement with the measurement, and thus we conclude that our Hamiltonian in Eq.~\ref{eq:lambda_ham} is a good description for the dynamics of the system.

The value we get is $\epsilon = 2\pi\times 50.8 ~\mathrm{MHz}$, and from it we obtain the effective Rabi rate of our oscillations $\Omega_\mathrm{R} = \frac{2g_3 \epsilon}{\Delta_r} = 2\pi\times 2 ~\mathrm{MHz}$. Recall that in Appendix C we estimated that for the same drive amplitude $g_3/\epsilon = 0.003$, and this is consistent with our measurements given that the direct resonator generator was set to be $-25~\mathrm{dB}$ lower than the nonlinear drive generator in our experiment.

\bibliographystyle{apsrev}
\bibliography{lambda_bib}

\begin{thebibliography}{44}
\expandafter\ifx\csname natexlab\endcsname\relax\def\natexlab#1{#1}\fi
\expandafter\ifx\csname bibnamefont\endcsname\relax
  \def\bibnamefont#1{#1}\fi
\expandafter\ifx\csname bibfnamefont\endcsname\relax
  \def\bibfnamefont#1{#1}\fi
\expandafter\ifx\csname citenamefont\endcsname\relax
  \def\citenamefont#1{#1}\fi
\expandafter\ifx\csname url\endcsname\relax
  \def\url#1{\texttt{#1}}\fi
\expandafter\ifx\csname urlprefix\endcsname\relax\def\urlprefix{URL }\fi
\providecommand{\bibinfo}[2]{#2}
\providecommand{\eprint}[2][]{\url{#2}}

\bibitem[{\citenamefont{Devoret and
  Schoelkopf}(2013)}]{devoret_superconducting_2013}
\bibinfo{author}{\bibfnamefont{M.~H.} \bibnamefont{Devoret}} \bibnamefont{and}
  \bibinfo{author}{\bibfnamefont{R.~J.} \bibnamefont{Schoelkopf}},
  \bibinfo{journal}{Science} \textbf{\bibinfo{volume}{339}},
  \bibinfo{pages}{1169} (\bibinfo{year}{2013}).

\bibitem[{\citenamefont{Liu et~al.}(2005)\citenamefont{Liu, You, Wei, Sun, and
  Nori}}]{liu_optical_2005}
\bibinfo{author}{\bibfnamefont{Y.-x.} \bibnamefont{Liu}},
  \bibinfo{author}{\bibfnamefont{J.~Q.} \bibnamefont{You}},
  \bibinfo{author}{\bibfnamefont{L.~F.} \bibnamefont{Wei}},
  \bibinfo{author}{\bibfnamefont{C.~P.} \bibnamefont{Sun}}, \bibnamefont{and}
  \bibinfo{author}{\bibfnamefont{F.}~\bibnamefont{Nori}},
  \bibinfo{journal}{Physical Review Letters} \textbf{\bibinfo{volume}{95}},
  \bibinfo{pages}{087001} (\bibinfo{year}{2005}).

\bibitem[{\citenamefont{Deppe et~al.}(2008)\citenamefont{Deppe, Mariantoni,
  Menzel, Marx, Saito, Kakuyanagi, Tanaka, Meno, Semba, Takayanagi
  et~al.}}]{deppe_two-photon_2008}
\bibinfo{author}{\bibfnamefont{F.}~\bibnamefont{Deppe}},
  \bibinfo{author}{\bibfnamefont{M.}~\bibnamefont{Mariantoni}},
  \bibinfo{author}{\bibfnamefont{E.~P.} \bibnamefont{Menzel}},
  \bibinfo{author}{\bibfnamefont{A.}~\bibnamefont{Marx}},
  \bibinfo{author}{\bibfnamefont{S.}~\bibnamefont{Saito}},
  \bibinfo{author}{\bibfnamefont{K.}~\bibnamefont{Kakuyanagi}},
  \bibinfo{author}{\bibfnamefont{H.}~\bibnamefont{Tanaka}},
  \bibinfo{author}{\bibfnamefont{T.}~\bibnamefont{Meno}},
  \bibinfo{author}{\bibfnamefont{K.}~\bibnamefont{Semba}},
  \bibinfo{author}{\bibfnamefont{H.}~\bibnamefont{Takayanagi}},
  \bibnamefont{et~al.}, \bibinfo{journal}{Nature Physics}
  \textbf{\bibinfo{volume}{4}}, \bibinfo{pages}{686} (\bibinfo{year}{2008}).

\bibitem[{\citenamefont{Harrabi et~al.}(2009)\citenamefont{Harrabi, Yoshihara,
  Niskanen, Nakamura, and Tsai}}]{harrabi_engineered_2009}
\bibinfo{author}{\bibfnamefont{K.}~\bibnamefont{Harrabi}},
  \bibinfo{author}{\bibfnamefont{F.}~\bibnamefont{Yoshihara}},
  \bibinfo{author}{\bibfnamefont{A.~O.} \bibnamefont{Niskanen}},
  \bibinfo{author}{\bibfnamefont{Y.}~\bibnamefont{Nakamura}}, \bibnamefont{and}
  \bibinfo{author}{\bibfnamefont{J.~S.} \bibnamefont{Tsai}},
  \bibinfo{journal}{Physical Review B} \textbf{\bibinfo{volume}{79}},
  \bibinfo{pages}{020507} (\bibinfo{year}{2009}).

\bibitem[{\citenamefont{Forn-Díaz et~al.}(2016)\citenamefont{Forn-Díaz,
  Romero, Harmans, Solano, and Mooij}}]{forn-diaz_broken_2016}
\bibinfo{author}{\bibfnamefont{P.}~\bibnamefont{Forn-Díaz}},
  \bibinfo{author}{\bibfnamefont{G.}~\bibnamefont{Romero}},
  \bibinfo{author}{\bibfnamefont{C.~J. P.~M.} \bibnamefont{Harmans}},
  \bibinfo{author}{\bibfnamefont{E.}~\bibnamefont{Solano}}, \bibnamefont{and}
  \bibinfo{author}{\bibfnamefont{J.~E.} \bibnamefont{Mooij}},
  \bibinfo{journal}{Scientific Reports} \textbf{\bibinfo{volume}{6}},
  \bibinfo{pages}{26720} (\bibinfo{year}{2016}).

\bibitem[{\citenamefont{Goetz et~al.}(2017)\citenamefont{Goetz, Deppe, Fedorov,
  Eder, Fischer, Pogorzalek, Xie, Marx, and
  Gross}}]{goetz_parity-engineered_2017}
\bibinfo{author}{\bibfnamefont{J.}~\bibnamefont{Goetz}},
  \bibinfo{author}{\bibfnamefont{F.}~\bibnamefont{Deppe}},
  \bibinfo{author}{\bibfnamefont{K.~G.} \bibnamefont{Fedorov}},
  \bibinfo{author}{\bibfnamefont{P.}~\bibnamefont{Eder}},
  \bibinfo{author}{\bibfnamefont{M.}~\bibnamefont{Fischer}},
  \bibinfo{author}{\bibfnamefont{S.}~\bibnamefont{Pogorzalek}},
  \bibinfo{author}{\bibfnamefont{E.}~\bibnamefont{Xie}},
  \bibinfo{author}{\bibfnamefont{A.}~\bibnamefont{Marx}}, \bibnamefont{and}
  \bibinfo{author}{\bibfnamefont{R.}~\bibnamefont{Gross}},
  \bibinfo{journal}{arXiv:1708.06405}  (\bibinfo{year}{2017}).

\bibitem[{\citenamefont{Lin et~al.}(2017)\citenamefont{Lin, Nguyen, Grabon,
  Miguel, Pankratova, and Manucharyan}}]{lin_protecting_2017}
\bibinfo{author}{\bibfnamefont{Y.-H.} \bibnamefont{Lin}},
  \bibinfo{author}{\bibfnamefont{L.~B.} \bibnamefont{Nguyen}},
  \bibinfo{author}{\bibfnamefont{N.}~\bibnamefont{Grabon}},
  \bibinfo{author}{\bibfnamefont{J.~S.} \bibnamefont{Miguel}},
  \bibinfo{author}{\bibfnamefont{N.}~\bibnamefont{Pankratova}},
  \bibnamefont{and} \bibinfo{author}{\bibfnamefont{V.~E.}
  \bibnamefont{Manucharyan}}, \bibinfo{journal}{arXiv:1705.07873}
  (\bibinfo{year}{2017}).

\bibitem[{\citenamefont{Blais et~al.}(2007)\citenamefont{Blais, Gambetta,
  Wallraff, Schuster, Girvin, Devoret, and
  Schoelkopf}}]{blais_quantum-information_2007}
\bibinfo{author}{\bibfnamefont{A.}~\bibnamefont{Blais}},
  \bibinfo{author}{\bibfnamefont{J.}~\bibnamefont{Gambetta}},
  \bibinfo{author}{\bibfnamefont{A.}~\bibnamefont{Wallraff}},
  \bibinfo{author}{\bibfnamefont{D.~I.} \bibnamefont{Schuster}},
  \bibinfo{author}{\bibfnamefont{S.~M.} \bibnamefont{Girvin}},
  \bibinfo{author}{\bibfnamefont{M.~H.} \bibnamefont{Devoret}},
  \bibnamefont{and} \bibinfo{author}{\bibfnamefont{R.~J.}
  \bibnamefont{Schoelkopf}}, \bibinfo{journal}{Physical Review A}
  \textbf{\bibinfo{volume}{75}}, \bibinfo{pages}{032329}
  (\bibinfo{year}{2007}).

\bibitem[{\citenamefont{Ithier et~al.}(2005)\citenamefont{Ithier, Collin,
  Joyez, Meeson, Vion, Esteve, Chiarello, Shnirman, Makhlin, Schriefl
  et~al.}}]{ithier_decoherence_2005}
\bibinfo{author}{\bibfnamefont{G.}~\bibnamefont{Ithier}},
  \bibinfo{author}{\bibfnamefont{E.}~\bibnamefont{Collin}},
  \bibinfo{author}{\bibfnamefont{P.}~\bibnamefont{Joyez}},
  \bibinfo{author}{\bibfnamefont{P.~J.} \bibnamefont{Meeson}},
  \bibinfo{author}{\bibfnamefont{D.}~\bibnamefont{Vion}},
  \bibinfo{author}{\bibfnamefont{D.}~\bibnamefont{Esteve}},
  \bibinfo{author}{\bibfnamefont{F.}~\bibnamefont{Chiarello}},
  \bibinfo{author}{\bibfnamefont{A.}~\bibnamefont{Shnirman}},
  \bibinfo{author}{\bibfnamefont{Y.}~\bibnamefont{Makhlin}},
  \bibinfo{author}{\bibfnamefont{J.}~\bibnamefont{Schriefl}},
  \bibnamefont{et~al.}, \bibinfo{journal}{Physical Review B}
  \textbf{\bibinfo{volume}{72}}, \bibinfo{pages}{134519}
  (\bibinfo{year}{2005}).

\bibitem[{\citenamefont{Doucot and Ioffe}(2012)}]{doucot_physical_2012}
\bibinfo{author}{\bibfnamefont{B.}~\bibnamefont{Doucot}} \bibnamefont{and}
  \bibinfo{author}{\bibfnamefont{L.~B.} \bibnamefont{Ioffe}},
  \bibinfo{journal}{Reports on Progress in Physics}
  \textbf{\bibinfo{volume}{75}}, \bibinfo{pages}{072001}
  (\bibinfo{year}{2012}).

\bibitem[{\citenamefont{Brooks et~al.}(2013)\citenamefont{Brooks, Kitaev, and
  Preskill}}]{brooks_protected_2013}
\bibinfo{author}{\bibfnamefont{P.}~\bibnamefont{Brooks}},
  \bibinfo{author}{\bibfnamefont{A.}~\bibnamefont{Kitaev}}, \bibnamefont{and}
  \bibinfo{author}{\bibfnamefont{J.}~\bibnamefont{Preskill}},
  \bibinfo{journal}{Phys. Rev. A} \textbf{\bibinfo{volume}{87}},
  \bibinfo{pages}{052306} (\bibinfo{year}{2013}).

\bibitem[{\citenamefont{Bell et~al.}(2014)\citenamefont{Bell, Paramanandam,
  Ioffe, and Gershenson}}]{bell_protected_2014}
\bibinfo{author}{\bibfnamefont{M.~T.} \bibnamefont{Bell}},
  \bibinfo{author}{\bibfnamefont{J.}~\bibnamefont{Paramanandam}},
  \bibinfo{author}{\bibfnamefont{L.~B.} \bibnamefont{Ioffe}}, \bibnamefont{and}
  \bibinfo{author}{\bibfnamefont{M.~E.} \bibnamefont{Gershenson}},
  \bibinfo{journal}{Physical Review Letters} \textbf{\bibinfo{volume}{112}},
  \bibinfo{pages}{167001} (\bibinfo{year}{2014}).

\bibitem[{\citenamefont{Kou et~al.}(2017{\natexlab{a}})\citenamefont{Kou,
  Smith, Vool, Brierley, Meier, Frunzio, Girvin, Glazman, and
  H.}}]{kou_fluxonium-based_2017}
\bibinfo{author}{\bibfnamefont{A.}~\bibnamefont{Kou}},
  \bibinfo{author}{\bibfnamefont{W.}~\bibnamefont{Smith}},
  \bibinfo{author}{\bibfnamefont{U.}~\bibnamefont{Vool}},
  \bibinfo{author}{\bibfnamefont{R.}~\bibnamefont{Brierley}},
  \bibinfo{author}{\bibfnamefont{H.}~\bibnamefont{Meier}},
  \bibinfo{author}{\bibfnamefont{L.}~\bibnamefont{Frunzio}},
  \bibinfo{author}{\bibfnamefont{S.}~\bibnamefont{Girvin}},
  \bibinfo{author}{\bibfnamefont{L.}~\bibnamefont{Glazman}}, \bibnamefont{and}
  \bibinfo{author}{\bibfnamefont{D.~M.} \bibnamefont{H.}},
  \bibinfo{journal}{Physical Review X} \textbf{\bibinfo{volume}{7}},
  \bibinfo{pages}{031037} (\bibinfo{year}{2017}{\natexlab{a}}).

\bibitem[{\citenamefont{Manucharyan et~al.}(2009)\citenamefont{Manucharyan,
  Koch, Glazman, and Devoret}}]{manucharyan_fluxonium:_2009}
\bibinfo{author}{\bibfnamefont{V.~E.} \bibnamefont{Manucharyan}},
  \bibinfo{author}{\bibfnamefont{J.}~\bibnamefont{Koch}},
  \bibinfo{author}{\bibfnamefont{L.~I.} \bibnamefont{Glazman}},
  \bibnamefont{and} \bibinfo{author}{\bibfnamefont{M.~H.}
  \bibnamefont{Devoret}}, \bibinfo{journal}{Science}
  \textbf{\bibinfo{volume}{326}}, \bibinfo{pages}{113} (\bibinfo{year}{2009}).

\bibitem[{\citenamefont{Manucharyan}(2011)}]{manucharyan_superinductance_2011}
\bibinfo{author}{\bibfnamefont{V.~E.} \bibnamefont{Manucharyan}}, Ph.D. thesis,
  \bibinfo{school}{Yale University} (\bibinfo{year}{2011}).

\bibitem[{\citenamefont{Pop et~al.}(2014)\citenamefont{Pop, Geerlings,
  Catelani, Schoelkopf, Glazman, and Devoret}}]{pop_coherent_2014}
\bibinfo{author}{\bibfnamefont{I.~M.} \bibnamefont{Pop}},
  \bibinfo{author}{\bibfnamefont{K.}~\bibnamefont{Geerlings}},
  \bibinfo{author}{\bibfnamefont{G.}~\bibnamefont{Catelani}},
  \bibinfo{author}{\bibfnamefont{R.~J.} \bibnamefont{Schoelkopf}},
  \bibinfo{author}{\bibfnamefont{L.~I.} \bibnamefont{Glazman}},
  \bibnamefont{and} \bibinfo{author}{\bibfnamefont{M.~H.}
  \bibnamefont{Devoret}}, \bibinfo{journal}{Nature}
  \textbf{\bibinfo{volume}{508}}, \bibinfo{pages}{369} (\bibinfo{year}{2014}).

\bibitem[{\citenamefont{Vool et~al.}(2014)\citenamefont{Vool, Pop, Sliwa, Abdo,
  Wang, Brecht, Gao, Shankar, Hatridge, Catelani
  et~al.}}]{vool_non-poissonian_2014}
\bibinfo{author}{\bibfnamefont{U.}~\bibnamefont{Vool}},
  \bibinfo{author}{\bibfnamefont{I.~M.} \bibnamefont{Pop}},
  \bibinfo{author}{\bibfnamefont{K.}~\bibnamefont{Sliwa}},
  \bibinfo{author}{\bibfnamefont{B.}~\bibnamefont{Abdo}},
  \bibinfo{author}{\bibfnamefont{C.}~\bibnamefont{Wang}},
  \bibinfo{author}{\bibfnamefont{T.}~\bibnamefont{Brecht}},
  \bibinfo{author}{\bibfnamefont{Y.~Y.} \bibnamefont{Gao}},
  \bibinfo{author}{\bibfnamefont{S.}~\bibnamefont{Shankar}},
  \bibinfo{author}{\bibfnamefont{M.}~\bibnamefont{Hatridge}},
  \bibinfo{author}{\bibfnamefont{G.}~\bibnamefont{Catelani}},
  \bibnamefont{et~al.}, \bibinfo{journal}{Phys. Rev. Lett.}
  \textbf{\bibinfo{volume}{113}}, \bibinfo{pages}{247001}
  (\bibinfo{year}{2014}).

\bibitem[{\citenamefont{Vool}(2017)}]{vool_engineering_2017}
\bibinfo{author}{\bibfnamefont{U.}~\bibnamefont{Vool}}, Ph.D. thesis,
  \bibinfo{school}{Yale University} (\bibinfo{year}{2017}).

\bibitem[{\citenamefont{Liu et~al.}(2006)\citenamefont{Liu, Wei, Tsai, and
  Nori}}]{liu_controllable_2006}
\bibinfo{author}{\bibfnamefont{Y.-x.} \bibnamefont{Liu}},
  \bibinfo{author}{\bibfnamefont{L.~F.} \bibnamefont{Wei}},
  \bibinfo{author}{\bibfnamefont{J.~S.} \bibnamefont{Tsai}}, \bibnamefont{and}
  \bibinfo{author}{\bibfnamefont{F.}~\bibnamefont{Nori}},
  \bibinfo{journal}{Physical Review Letters} \textbf{\bibinfo{volume}{96}},
  \bibinfo{pages}{067003} (\bibinfo{year}{2006}).

\bibitem[{\citenamefont{Frattini et~al.}(2017)\citenamefont{Frattini, Vool,
  Shankar, Narla, Sliwa, and Devoret}}]{frattini_3-wave_2017}
\bibinfo{author}{\bibfnamefont{N.~E.} \bibnamefont{Frattini}},
  \bibinfo{author}{\bibfnamefont{U.}~\bibnamefont{Vool}},
  \bibinfo{author}{\bibfnamefont{S.}~\bibnamefont{Shankar}},
  \bibinfo{author}{\bibfnamefont{A.}~\bibnamefont{Narla}},
  \bibinfo{author}{\bibfnamefont{K.~M.} \bibnamefont{Sliwa}}, \bibnamefont{and}
  \bibinfo{author}{\bibfnamefont{M.~H.} \bibnamefont{Devoret}},
  \bibinfo{journal}{Applied Physics Letters} \textbf{\bibinfo{volume}{110}},
  \bibinfo{pages}{222603} (\bibinfo{year}{2017}).

\bibitem[{\citenamefont{Mooij et~al.}(1999)\citenamefont{Mooij, Orlando,
  Levitov, Tian, Wal, and Lloyd}}]{mooij_josephson_1999}
\bibinfo{author}{\bibfnamefont{J.~E.} \bibnamefont{Mooij}},
  \bibinfo{author}{\bibfnamefont{T.~P.} \bibnamefont{Orlando}},
  \bibinfo{author}{\bibfnamefont{L.}~\bibnamefont{Levitov}},
  \bibinfo{author}{\bibfnamefont{L.}~\bibnamefont{Tian}},
  \bibinfo{author}{\bibfnamefont{C.~H. v.~d.} \bibnamefont{Wal}},
  \bibnamefont{and} \bibinfo{author}{\bibfnamefont{S.}~\bibnamefont{Lloyd}},
  \bibinfo{journal}{Science} \textbf{\bibinfo{volume}{285}},
  \bibinfo{pages}{1036} (\bibinfo{year}{1999}).

\bibitem[{\citenamefont{Wal et~al.}(2000)\citenamefont{Wal, Haar, Wilhelm,
  Schouten, Harmans, Orlando, Lloyd, and Mooij}}]{wal_quantum_2000}
\bibinfo{author}{\bibfnamefont{C.~H. v.~d.} \bibnamefont{Wal}},
  \bibinfo{author}{\bibfnamefont{A.~C. J.~t.} \bibnamefont{Haar}},
  \bibinfo{author}{\bibfnamefont{F.~K.} \bibnamefont{Wilhelm}},
  \bibinfo{author}{\bibfnamefont{R.~N.} \bibnamefont{Schouten}},
  \bibinfo{author}{\bibfnamefont{C.~J. P.~M.} \bibnamefont{Harmans}},
  \bibinfo{author}{\bibfnamefont{T.~P.} \bibnamefont{Orlando}},
  \bibinfo{author}{\bibfnamefont{S.}~\bibnamefont{Lloyd}}, \bibnamefont{and}
  \bibinfo{author}{\bibfnamefont{J.~E.} \bibnamefont{Mooij}},
  \bibinfo{journal}{Science} \textbf{\bibinfo{volume}{290}},
  \bibinfo{pages}{773} (\bibinfo{year}{2000}).

\bibitem[{\citenamefont{Zorin}(2016)}]{zorin_josephson_2016}
\bibinfo{author}{\bibfnamefont{A.~B.} \bibnamefont{Zorin}},
  \bibinfo{journal}{Physical Review Applied} \textbf{\bibinfo{volume}{6}},
  \bibinfo{pages}{034006} (\bibinfo{year}{2016}).

\bibitem[{\citenamefont{Zorin et~al.}(2017)\citenamefont{Zorin, Khabipov,
  Dietel, and Dolata}}]{zorin_traveling-wave_2017}
\bibinfo{author}{\bibfnamefont{A.~B.} \bibnamefont{Zorin}},
  \bibinfo{author}{\bibfnamefont{M.}~\bibnamefont{Khabipov}},
  \bibinfo{author}{\bibfnamefont{J.}~\bibnamefont{Dietel}}, \bibnamefont{and}
  \bibinfo{author}{\bibfnamefont{R.}~\bibnamefont{Dolata}},
  \bibinfo{journal}{arXiv:1705.02859}  (\bibinfo{year}{2017}).

\bibitem[{\citenamefont{Lecocq et~al.}(2011)\citenamefont{Lecocq, Pop, Peng,
  Matei, Crozes, Fournier, {Cécile Naud}, Guichard, and
  Buisson}}]{lecocq_junction_2011}
\bibinfo{author}{\bibfnamefont{F.}~\bibnamefont{Lecocq}},
  \bibinfo{author}{\bibfnamefont{I.~M.} \bibnamefont{Pop}},
  \bibinfo{author}{\bibfnamefont{Z.}~\bibnamefont{Peng}},
  \bibinfo{author}{\bibfnamefont{I.}~\bibnamefont{Matei}},
  \bibinfo{author}{\bibfnamefont{T.}~\bibnamefont{Crozes}},
  \bibinfo{author}{\bibfnamefont{T.}~\bibnamefont{Fournier}},
  \bibinfo{author}{\bibnamefont{{Cécile Naud}}},
  \bibinfo{author}{\bibfnamefont{W.}~\bibnamefont{Guichard}}, \bibnamefont{and}
  \bibinfo{author}{\bibfnamefont{O.}~\bibnamefont{Buisson}},
  \bibinfo{journal}{Nanotechnology} \textbf{\bibinfo{volume}{22}},
  \bibinfo{pages}{315302} (\bibinfo{year}{2011}).

\bibitem[{\citenamefont{Pop et~al.}(2012)\citenamefont{Pop, Fournier, Crozes,
  Lecocq, Matei, Pannetier, Buisson, and Guichard}}]{pop_fabrication_2012}
\bibinfo{author}{\bibfnamefont{I.~M.} \bibnamefont{Pop}},
  \bibinfo{author}{\bibfnamefont{T.}~\bibnamefont{Fournier}},
  \bibinfo{author}{\bibfnamefont{T.}~\bibnamefont{Crozes}},
  \bibinfo{author}{\bibfnamefont{F.}~\bibnamefont{Lecocq}},
  \bibinfo{author}{\bibfnamefont{I.}~\bibnamefont{Matei}},
  \bibinfo{author}{\bibfnamefont{B.}~\bibnamefont{Pannetier}},
  \bibinfo{author}{\bibfnamefont{O.}~\bibnamefont{Buisson}}, \bibnamefont{and}
  \bibinfo{author}{\bibfnamefont{W.}~\bibnamefont{Guichard}},
  \bibinfo{journal}{J. Vac. Sci. Technol. B} \textbf{\bibinfo{volume}{30}},
  \bibinfo{pages}{010607} (\bibinfo{year}{2012}).

\bibitem[{\citenamefont{Kou et~al.}(2017{\natexlab{b}})\citenamefont{Kou,
  Smith, Vool, Pop, Sliwa, Hatridge, Frunzio, and
  Devoret}}]{kou_simultaneous_2017}
\bibinfo{author}{\bibfnamefont{A.}~\bibnamefont{Kou}},
  \bibinfo{author}{\bibfnamefont{W.~C.} \bibnamefont{Smith}},
  \bibinfo{author}{\bibfnamefont{U.}~\bibnamefont{Vool}},
  \bibinfo{author}{\bibfnamefont{I.~M.} \bibnamefont{Pop}},
  \bibinfo{author}{\bibfnamefont{K.~M.} \bibnamefont{Sliwa}},
  \bibinfo{author}{\bibfnamefont{M.~H.} \bibnamefont{Hatridge}},
  \bibinfo{author}{\bibfnamefont{L.}~\bibnamefont{Frunzio}}, \bibnamefont{and}
  \bibinfo{author}{\bibfnamefont{M.~H.} \bibnamefont{Devoret}},
  \bibinfo{journal}{arXiv:1705.05712}  (\bibinfo{year}{2017}{\natexlab{b}}).

\bibitem[{\citenamefont{Yang et~al.}(2004)\citenamefont{Yang, Chu, and
  Han}}]{yang_quantum_2004}
\bibinfo{author}{\bibfnamefont{C.-P.} \bibnamefont{Yang}},
  \bibinfo{author}{\bibfnamefont{S.-I.} \bibnamefont{Chu}}, \bibnamefont{and}
  \bibinfo{author}{\bibfnamefont{S.}~\bibnamefont{Han}},
  \bibinfo{journal}{Physical Review Letters} \textbf{\bibinfo{volume}{92}},
  \bibinfo{pages}{117902} (\bibinfo{year}{2004}).

\bibitem[{\citenamefont{Murali et~al.}(2004)\citenamefont{Murali, Dutton,
  Oliver, Crankshaw, and Orlando}}]{murali_probing_2004}
\bibinfo{author}{\bibfnamefont{K.~V. R.~M.} \bibnamefont{Murali}},
  \bibinfo{author}{\bibfnamefont{Z.}~\bibnamefont{Dutton}},
  \bibinfo{author}{\bibfnamefont{W.~D.} \bibnamefont{Oliver}},
  \bibinfo{author}{\bibfnamefont{D.~S.} \bibnamefont{Crankshaw}},
  \bibnamefont{and} \bibinfo{author}{\bibfnamefont{T.~P.}
  \bibnamefont{Orlando}}, \bibinfo{journal}{Physical Review Letters}
  \textbf{\bibinfo{volume}{93}}, \bibinfo{pages}{087003}
  (\bibinfo{year}{2004}).

\bibitem[{\citenamefont{Valenzuela et~al.}(2006)\citenamefont{Valenzuela,
  Oliver, Berns, Berggren, Levitov, and
  Orlando}}]{valenzuela_microwave-induced_2006}
\bibinfo{author}{\bibfnamefont{S.~O.} \bibnamefont{Valenzuela}},
  \bibinfo{author}{\bibfnamefont{W.~D.} \bibnamefont{Oliver}},
  \bibinfo{author}{\bibfnamefont{D.~M.} \bibnamefont{Berns}},
  \bibinfo{author}{\bibfnamefont{K.~K.} \bibnamefont{Berggren}},
  \bibinfo{author}{\bibfnamefont{L.~S.} \bibnamefont{Levitov}},
  \bibnamefont{and} \bibinfo{author}{\bibfnamefont{T.~P.}
  \bibnamefont{Orlando}}, \bibinfo{journal}{Science}
  \textbf{\bibinfo{volume}{314}}, \bibinfo{pages}{1589} (\bibinfo{year}{2006}).

\bibitem[{\citenamefont{Grajcar et~al.}(2008)\citenamefont{Grajcar, van~der
  Ploeg, Izmalkov, Il’ichev, Meyer, Fedorov, Shnirman, and
  Sch{\"o}n}}]{grajcar_sisyphus_2008}
\bibinfo{author}{\bibfnamefont{M.}~\bibnamefont{Grajcar}},
  \bibinfo{author}{\bibfnamefont{S.~H.~W.} \bibnamefont{van~der Ploeg}},
  \bibinfo{author}{\bibfnamefont{A.}~\bibnamefont{Izmalkov}},
  \bibinfo{author}{\bibfnamefont{E.}~\bibnamefont{Il’ichev}},
  \bibinfo{author}{\bibfnamefont{H.-G.} \bibnamefont{Meyer}},
  \bibinfo{author}{\bibfnamefont{A.}~\bibnamefont{Fedorov}},
  \bibinfo{author}{\bibfnamefont{A.}~\bibnamefont{Shnirman}}, \bibnamefont{and}
  \bibinfo{author}{\bibfnamefont{G.}~\bibnamefont{Sch{\"o}n}},
  \bibinfo{journal}{Nature Physics} \textbf{\bibinfo{volume}{4}},
  \bibinfo{pages}{612} (\bibinfo{year}{2008}).

\bibitem[{\citenamefont{Inomata et~al.}(2014)\citenamefont{Inomata, Koshino,
  Lin, Oliver, Tsai, Nakamura, and Yamamoto}}]{inomata_microwave_2014}
\bibinfo{author}{\bibfnamefont{K.}~\bibnamefont{Inomata}},
  \bibinfo{author}{\bibfnamefont{K.}~\bibnamefont{Koshino}},
  \bibinfo{author}{\bibfnamefont{Z.~R.} \bibnamefont{Lin}},
  \bibinfo{author}{\bibfnamefont{W.~D.} \bibnamefont{Oliver}},
  \bibinfo{author}{\bibfnamefont{J.~S.} \bibnamefont{Tsai}},
  \bibinfo{author}{\bibfnamefont{Y.}~\bibnamefont{Nakamura}}, \bibnamefont{and}
  \bibinfo{author}{\bibfnamefont{T.}~\bibnamefont{Yamamoto}},
  \bibinfo{journal}{Physical Review Letters} \textbf{\bibinfo{volume}{113}},
  \bibinfo{pages}{063604} (\bibinfo{year}{2014}).

\bibitem[{\citenamefont{Inomata et~al.}(2016)\citenamefont{Inomata, Lin,
  Koshino, Oliver, Tsai, Yamamoto, and Nakamura}}]{inomata_single_2016}
\bibinfo{author}{\bibfnamefont{K.}~\bibnamefont{Inomata}},
  \bibinfo{author}{\bibfnamefont{Z.}~\bibnamefont{Lin}},
  \bibinfo{author}{\bibfnamefont{K.}~\bibnamefont{Koshino}},
  \bibinfo{author}{\bibfnamefont{W.~D.} \bibnamefont{Oliver}},
  \bibinfo{author}{\bibfnamefont{J.-S.} \bibnamefont{Tsai}},
  \bibinfo{author}{\bibfnamefont{T.}~\bibnamefont{Yamamoto}}, \bibnamefont{and}
  \bibinfo{author}{\bibfnamefont{Y.}~\bibnamefont{Nakamura}},
  \bibinfo{journal}{Nature Communications} \textbf{\bibinfo{volume}{7}},
  \bibinfo{pages}{12303} (\bibinfo{year}{2016}).

\bibitem[{\citenamefont{Kelly et~al.}(2010)\citenamefont{Kelly, Dutton,
  Schlafer, Mookerji, Ohki, Kline, and Pappas}}]{kelly_direct_2010}
\bibinfo{author}{\bibfnamefont{W.~R.} \bibnamefont{Kelly}},
  \bibinfo{author}{\bibfnamefont{Z.}~\bibnamefont{Dutton}},
  \bibinfo{author}{\bibfnamefont{J.}~\bibnamefont{Schlafer}},
  \bibinfo{author}{\bibfnamefont{B.}~\bibnamefont{Mookerji}},
  \bibinfo{author}{\bibfnamefont{T.~A.} \bibnamefont{Ohki}},
  \bibinfo{author}{\bibfnamefont{J.~S.} \bibnamefont{Kline}}, \bibnamefont{and}
  \bibinfo{author}{\bibfnamefont{D.~P.} \bibnamefont{Pappas}},
  \bibinfo{journal}{Physical Review Letters} \textbf{\bibinfo{volume}{104}},
  \bibinfo{pages}{163601} (\bibinfo{year}{2010}).

\bibitem[{\citenamefont{Novikov et~al.}(2016)\citenamefont{Novikov, Sweeney,
  Robinson, Premaratne, Suri, Wellstood, and Palmer}}]{novikov_raman_2016}
\bibinfo{author}{\bibfnamefont{S.}~\bibnamefont{Novikov}},
  \bibinfo{author}{\bibfnamefont{T.}~\bibnamefont{Sweeney}},
  \bibinfo{author}{\bibfnamefont{J.~E.} \bibnamefont{Robinson}},
  \bibinfo{author}{\bibfnamefont{S.~P.} \bibnamefont{Premaratne}},
  \bibinfo{author}{\bibfnamefont{B.}~\bibnamefont{Suri}},
  \bibinfo{author}{\bibfnamefont{F.~C.} \bibnamefont{Wellstood}},
  \bibnamefont{and} \bibinfo{author}{\bibfnamefont{B.~S.}
  \bibnamefont{Palmer}}, \bibinfo{journal}{Nature Physics}
  \textbf{\bibinfo{volume}{12}}, \bibinfo{pages}{75} (\bibinfo{year}{2016}).

\bibitem[{\citenamefont{Earnest et~al.}(2017)\citenamefont{Earnest, Chakram,
  Lu, Irons, Naik, Leung, Lawrence, Koch, and
  Schuster}}]{earnest_realization_2017}
\bibinfo{author}{\bibfnamefont{N.}~\bibnamefont{Earnest}},
  \bibinfo{author}{\bibfnamefont{S.}~\bibnamefont{Chakram}},
  \bibinfo{author}{\bibfnamefont{Y.}~\bibnamefont{Lu}},
  \bibinfo{author}{\bibfnamefont{N.}~\bibnamefont{Irons}},
  \bibinfo{author}{\bibfnamefont{R.~K.} \bibnamefont{Naik}},
  \bibinfo{author}{\bibfnamefont{N.}~\bibnamefont{Leung}},
  \bibinfo{author}{\bibfnamefont{J.}~\bibnamefont{Lawrence}},
  \bibinfo{author}{\bibfnamefont{J.}~\bibnamefont{Koch}}, \bibnamefont{and}
  \bibinfo{author}{\bibfnamefont{D.~I.} \bibnamefont{Schuster}},
  \bibinfo{journal}{arXiv:1707.00656}  (\bibinfo{year}{2017}).

\bibitem[{\citenamefont{Wallraff et~al.}(2007)\citenamefont{Wallraff, Schuster,
  Blais, Gambetta, Schreier, Frunzio, Devoret, Girvin, and
  Schoelkopf}}]{wallraff_sideband_2007}
\bibinfo{author}{\bibfnamefont{A.}~\bibnamefont{Wallraff}},
  \bibinfo{author}{\bibfnamefont{D.~I.} \bibnamefont{Schuster}},
  \bibinfo{author}{\bibfnamefont{A.}~\bibnamefont{Blais}},
  \bibinfo{author}{\bibfnamefont{J.~M.} \bibnamefont{Gambetta}},
  \bibinfo{author}{\bibfnamefont{J.}~\bibnamefont{Schreier}},
  \bibinfo{author}{\bibfnamefont{L.}~\bibnamefont{Frunzio}},
  \bibinfo{author}{\bibfnamefont{M.~H.} \bibnamefont{Devoret}},
  \bibinfo{author}{\bibfnamefont{S.~M.} \bibnamefont{Girvin}},
  \bibnamefont{and} \bibinfo{author}{\bibfnamefont{R.~J.}
  \bibnamefont{Schoelkopf}}, \bibinfo{journal}{Physical Review Letters}
  \textbf{\bibinfo{volume}{99}}, \bibinfo{pages}{050501}
  (\bibinfo{year}{2007}).

\bibitem[{\citenamefont{You and Nori}(2011)}]{you_atomic_2011}
\bibinfo{author}{\bibfnamefont{J.~Q.} \bibnamefont{You}} \bibnamefont{and}
  \bibinfo{author}{\bibfnamefont{F.}~\bibnamefont{Nori}},
  \bibinfo{journal}{Nature} \textbf{\bibinfo{volume}{474}},
  \bibinfo{pages}{589} (\bibinfo{year}{2011}).

\bibitem[{\citenamefont{Kerckhoff et~al.}(2010)\citenamefont{Kerckhoff, Nurdin,
  Pavlichin, and Mabuchi}}]{kerckhoff_designing_2010}
\bibinfo{author}{\bibfnamefont{J.}~\bibnamefont{Kerckhoff}},
  \bibinfo{author}{\bibfnamefont{H.~I.} \bibnamefont{Nurdin}},
  \bibinfo{author}{\bibfnamefont{D.~S.} \bibnamefont{Pavlichin}},
  \bibnamefont{and} \bibinfo{author}{\bibfnamefont{H.}~\bibnamefont{Mabuchi}},
  \bibinfo{journal}{Physical Review Letters} \textbf{\bibinfo{volume}{105}},
  \bibinfo{pages}{040502} (\bibinfo{year}{2010}).

\bibitem[{\citenamefont{Campagne-Ibarcq
  et~al.}(2017)\citenamefont{Campagne-Ibarcq, Zalys-Geller, Narla, Shankar,
  Reinhold, Burkhart, Axline, Pfaff, Frunzio, Schoelkopf
  et~al.}}]{campagne-ibarcq_deterministic_2017}
\bibinfo{author}{\bibfnamefont{P.}~\bibnamefont{Campagne-Ibarcq}},
  \bibinfo{author}{\bibfnamefont{E.}~\bibnamefont{Zalys-Geller}},
  \bibinfo{author}{\bibfnamefont{A.}~\bibnamefont{Narla}},
  \bibinfo{author}{\bibfnamefont{S.}~\bibnamefont{Shankar}},
  \bibinfo{author}{\bibfnamefont{P.}~\bibnamefont{Reinhold}},
  \bibinfo{author}{\bibfnamefont{L.~D.} \bibnamefont{Burkhart}},
  \bibinfo{author}{\bibfnamefont{C.~J.} \bibnamefont{Axline}},
  \bibinfo{author}{\bibfnamefont{W.}~\bibnamefont{Pfaff}},
  \bibinfo{author}{\bibfnamefont{L.}~\bibnamefont{Frunzio}},
  \bibinfo{author}{\bibfnamefont{R.~J.} \bibnamefont{Schoelkopf}},
  \bibnamefont{et~al.}, \bibinfo{journal}{arXiv:1712.05854}
  (\bibinfo{year}{2017}).

\bibitem[{\citenamefont{Axline et~al.}(2017)\citenamefont{Axline, Burkhart,
  Pfaff, Zhang, Chou, Campagne-Ibarcq, Reinhold, Frunzio, Girvin, Jiang
  et~al.}}]{axline_-demand_2017}
\bibinfo{author}{\bibfnamefont{C.}~\bibnamefont{Axline}},
  \bibinfo{author}{\bibfnamefont{L.}~\bibnamefont{Burkhart}},
  \bibinfo{author}{\bibfnamefont{W.}~\bibnamefont{Pfaff}},
  \bibinfo{author}{\bibfnamefont{M.}~\bibnamefont{Zhang}},
  \bibinfo{author}{\bibfnamefont{K.}~\bibnamefont{Chou}},
  \bibinfo{author}{\bibfnamefont{P.}~\bibnamefont{Campagne-Ibarcq}},
  \bibinfo{author}{\bibfnamefont{P.}~\bibnamefont{Reinhold}},
  \bibinfo{author}{\bibfnamefont{L.}~\bibnamefont{Frunzio}},
  \bibinfo{author}{\bibfnamefont{S.~M.} \bibnamefont{Girvin}},
  \bibinfo{author}{\bibfnamefont{L.}~\bibnamefont{Jiang}},
  \bibnamefont{et~al.}, \bibinfo{journal}{arXiv:1712.05832}
  (\bibinfo{year}{2017}).

\bibitem[{\citenamefont{Kurpiers et~al.}(2017)\citenamefont{Kurpiers, Magnard,
  Walter, Royer, Pechal, Heinsoo, Salathé, Akin, Storz, Besse
  et~al.}}]{kurpiers_deterministic_2017}
\bibinfo{author}{\bibfnamefont{P.}~\bibnamefont{Kurpiers}},
  \bibinfo{author}{\bibfnamefont{P.}~\bibnamefont{Magnard}},
  \bibinfo{author}{\bibfnamefont{T.}~\bibnamefont{Walter}},
  \bibinfo{author}{\bibfnamefont{B.}~\bibnamefont{Royer}},
  \bibinfo{author}{\bibfnamefont{M.}~\bibnamefont{Pechal}},
  \bibinfo{author}{\bibfnamefont{J.}~\bibnamefont{Heinsoo}},
  \bibinfo{author}{\bibfnamefont{Y.}~\bibnamefont{Salathé}},
  \bibinfo{author}{\bibfnamefont{A.}~\bibnamefont{Akin}},
  \bibinfo{author}{\bibfnamefont{S.}~\bibnamefont{Storz}},
  \bibinfo{author}{\bibfnamefont{J.-C.} \bibnamefont{Besse}},
  \bibnamefont{et~al.}, \bibinfo{journal}{arXiv:1712.08593}
  (\bibinfo{year}{2017}).

\bibitem[{\citenamefont{Vool and Devoret}(2017)}]{vool_introduction_2017}
\bibinfo{author}{\bibfnamefont{U.}~\bibnamefont{Vool}} \bibnamefont{and}
  \bibinfo{author}{\bibfnamefont{M.}~\bibnamefont{Devoret}},
  \bibinfo{journal}{International Journal of Circuit Theory and Applications}
  \textbf{\bibinfo{volume}{45}}, \bibinfo{pages}{897} (\bibinfo{year}{2017}).

\bibitem[{\citenamefont{Smith et~al.}(2016)\citenamefont{Smith, Kou, Vool, Pop,
  Frunzio, Schoelkopf, and Devoret}}]{smith_quantization_2016}
\bibinfo{author}{\bibfnamefont{W.~C.} \bibnamefont{Smith}},
  \bibinfo{author}{\bibfnamefont{A.}~\bibnamefont{Kou}},
  \bibinfo{author}{\bibfnamefont{U.}~\bibnamefont{Vool}},
  \bibinfo{author}{\bibfnamefont{I.~M.} \bibnamefont{Pop}},
  \bibinfo{author}{\bibfnamefont{L.}~\bibnamefont{Frunzio}},
  \bibinfo{author}{\bibfnamefont{R.~J.} \bibnamefont{Schoelkopf}},
  \bibnamefont{and} \bibinfo{author}{\bibfnamefont{M.~H.}
  \bibnamefont{Devoret}}, \bibinfo{journal}{Physical Review B}
  \textbf{\bibinfo{volume}{94}}, \bibinfo{pages}{144507}
  (\bibinfo{year}{2016}).

\end{thebibliography}

\end{document}